\documentclass[]{elsarticle}

\makeatletter
\def\ps@pprintTitle{%
 \let\@oddhead\@empty
 \let\@evenhead\@empty
 \def\@oddfoot{}%
 \let\@evenfoot\@oddfoot}
\makeatother
\usepackage{url}

\usepackage{epsfig,latexsym,epsf,color}
\usepackage{graphicx,graphics, amsmath,amssymb,latexsym,epstopdf}
\usepackage{subfigure}
\usepackage{paralist}
\usepackage{bm}
\usepackage{algorithm}
\usepackage{algpseudocode}
\usepackage[only,tensf,]{rawfonts}

\newcommand{\tx}[1]{\text{tx}(l)}
\newcommand{\rx}[1]{\text{rx}(l)}

\newcommand{\ie}{{\it i.e.}}

\newcommand{\separator}{
  \begin{center}
    \rule{\columnwidth}{0.3mm}
  \end{center}
}

\newcommand{\bi}{\begin{itemize}}
\newcommand{\ei}{\end{itemize}}
\newcommand{\be}{\begin{enumerate}}
\newcommand{\ee}{\end{enumerate}}

\newcommand{\beq}{\begin{eqnarray*}}
\newcommand{\eeq}{\end{eqnarray*}}
\newcommand{\beqn}{\begin{eqnarray}}
\newcommand{\eeqn}{\end{eqnarray}}

\newcommand\etal{{\em et al.}}

\begin{document}

\begin{frontmatter}

\title{Sequential Movie Genre Prediction using Average Transition Probability with Clustering}

\author{Jihyeon Kim}
\ead{zizi39028@gmail.com}
\author{Jinkyung Kim}
\ead{lxsz987@gmail.com}

\author{Jaeyoung Choi\corref{cor1}}
\ead{jychoi19@gachon.ac.kr}

\cortext[cor1]{Corresponding author, Tel.: +82-31-750-5829}

\address{School of Computing, Gachon University, 1342 Seongnamdaero, Sujeong-gu, Seongnam-si, Gyeongi-do 13120}

\begin{abstract}
 In recent movie recommendations, predicting the user's sequential behavior and suggesting the next movie to watch is one of the most important issues. However, capturing such sequential behavior is not easy because each user's short-term or long-term behavior must be taken into account. For this reason, many research results show that the performance of recommending a specific movie is not very high in a sequential recommendation. In this paper, we propose a cluster-based method for classifying users with similar movie purchase patterns and a movie genre prediction algorithm rather than the movie itself considering their short-term and long-term behaviors. The movie genre prediction does not recommend a specific movie, but it predicts the genre for the next movie to watch in consideration of each user's preference for the movie genre based on the genre included in the movie. Through this, it is possible to provide appropriate guidelines for recommending movies including the genre to users who tend to prefer a specific genre. In particular, in this paper, users with similar genre preferences are organized into clusters to recommend genres, and in clusters that do not have relatively specific tendencies, genre prediction is performed by appropriately trimming genres that are not necessary for recommendation in order to improve performance.
We evaluate our method on well-known movie datasets, and qualitatively
that it captures personalized dynamics and is able to make
meaningful recommendations.
\end{abstract}


\end{frontmatter}

\section{Introduction}
\label{sec:intro}
One of the most important parts of a recommendation system is to model the interactions between
users and items, as well as the relationships amongst the
items themselves \cite{He2016}.
The former one is usually called item-to-user
recommendation and the latter one is called item-to-item recommendation.
In a sequential recommendation, user preferences and
sequential patterns can be extracted by the above two kinds of
interactions.
However, it is not an easy task to learn the personalized
sequential behavior from collaborative data since long and short term
dynamics of the users have to be carefully considered
for both personalization and sequential transitions.
Especially, if the data is sparse, estimating parameters of learning models becomes very hard.
For this reason, a main object is to design
personalized models of user behavior using the
users’ recent purchase histories for the sequential recommendation systems \cite{Kang2018}. To reflect the short-term dynamics of the user's behavior,
Markov Chain (MC) approaches have been used, which assume
that the next action depends on only the previous action. Since the last related
item is in general the key factor affecting the user’s next action, first-order MC based methods
show strong performance, especially on sparse datasets \cite{He2018}.
For the long-term dynamics of user's behavior, Recurrent Neural Networks (RNNs) have been used
to reflect all previous actions with a hidden state, which is
used to predict the next action \cite{Hidasi2018}.
Both approaches, while strong in specific cases, are quite
limited to certain types of data. MC-based approaches work well in high sparsity
settings, but may be difficult to obtain the intricate dynamics
of long-term complex scenarios. Different from this, RNNs work well in such long-term scenarios, which require large amounts of data.
\begin{figure*}[t!]
\begin{center} \centering
\includegraphics[width=1\linewidth]{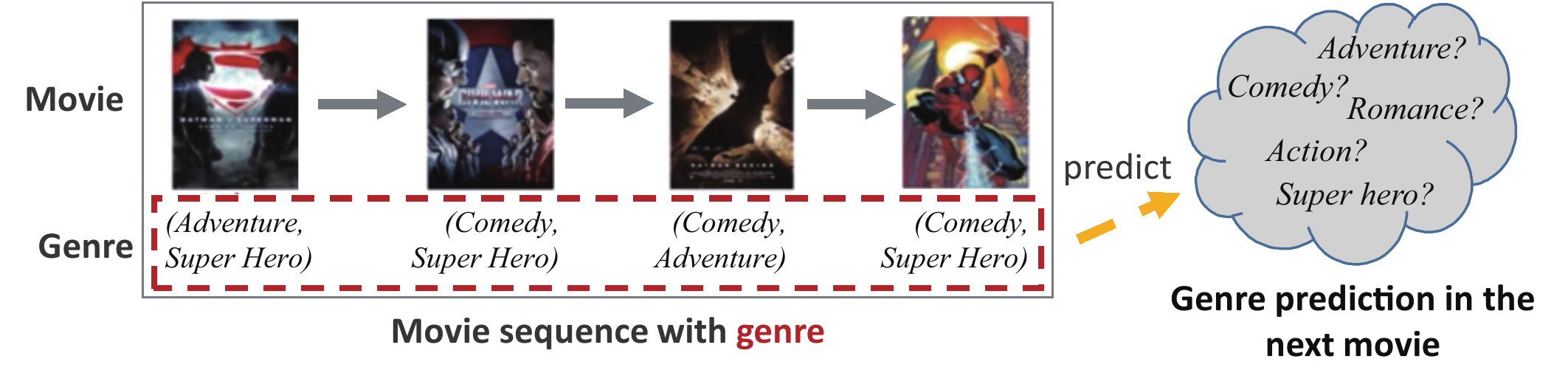}
\caption{Movie Genre Prediction.}
\label{fig:ER}
\vspace{-0.6cm}
\end{center}
\end{figure*}
Recently, session-based recommendation systems (SBRSs) have been investigated \cite{Wang2021} as a new approach to recommendation systems. Different from other Collaborative Filtering (CF)-based recommendation systems, which usually model long-term yet static
user preferences, SBRSs try to capture short-term but dynamic user preference for more sensitive accurate recommendations to the evolution of their session contexts.
Different from purchasing one specific item, the authors in  \cite{Rendle10} considered the sequential recommendation scenario that users purchase some similar items simultaneously. In this scenario, proposing each
user a personalized ``list of items" is the main object of the recommendation.
Similar to this approach, one can consider an attribute prediction for movie data to the users.
For instance, the
genre/category of a movie can be an important attribute for user/item similarity in the recommendation systems. Moreover, this information is provided when
new content is created. Based on this,
the authors in \cite{Choi2012} considered a movie recommendation system
based on genre correlations, to improve the existing
genre correlation algorithm, and compare the obtained results
using the previous algorithm and the proposed algorithm. In \cite{Kim2012}, the authors introduced a recommender system using movie genre similarity and preferred genres.
However, in these works, only a static situation was considered, not a sequentially changing recommendation systems.

In this paper, we first consider the prediction of movie genres included in preferred movies before recommending movies. The genre is one of the important features of a movie, which gives guidelines on which movies each user prefers. In the sequential movie recommendation system, we extract the genre included in the movie each user watched and studied the genre preference, and conducted a study on what movie genre the user will see next.

Our main contributions are described as follows:
\smallskip
\begin{itemize}

\item First, different from most prior researches of movie recommendation systems, we focus on the genres, which are included in a movie rather than the movie itself as a sequential prediction item. Although it cannot be used directly in sequential movie recommendation systems, it can show how well genre-based prediction works in learning user preferences.

\item Second, for predicting the long-term user's preference behavior of movie genres, we use RNN-based learning models that show the best performance recently. Further, we consider an Average Transition Probability (ATV) between genres as a Markov chain to reflect the short-term behavior of the user's preference as in \cite{Kang2018}. To see the effect of the average transition probability, we consider four kinds of training data with combining genre vectors.

\item Third, we propose a clustering approach based on the $k$-means clustering, which has similar preferences for movie genre. In order to improve the prediction performance, we also propose a method for properly trimming genres that act badly on performance using the results obtained based on the RNN-based sequential learning models presented above.

\item Finally, we evaluate
our method on well-known movie datasets, and qualitatively
that it captures personalized dynamics and is able to make
meaningful recommendations for the movie genres. The results show that clustering with trimming improves the prediction performance whereas applying ATV has a very negligible performance improvement.

\end{itemize}
The remainder of this paper is organized as follows. Section \ref{sec:related} discusses related studies. In Section
\ref{sec:method}, our clustering and training methods are presented. In Section \ref{sec:simulation}, the experiment results of our proposed methods are presented and some limitation and future works are given in Section \ref{sec:discussion}. In Section \ref{sec:conclusion}, we conclude the paper.

\section{Related Works}
\label{sec:related}
In recent sequential recommendation systems, most works focus on how to predict the short-term and long-term preference dynamic of users. As a short-term dynamic, the Markov chain approach has been studied.
Zimdars \etal \cite{Zimdars01} described
a sequential recommender based on the Markov chains. They
studied how to catch sequential patterns to predict the
next state with a standard predictor such as a decision tree.
Rendle \etal \cite{Rendle10} proposed a Factorizing Personalized Markov Chain (FPMC), where they modeled based on user-specific transition with Markov Chain, about the history of a basket. FPMC propagates information among users, items, transitions which has similar favor or patterns to extract the sequential pattern.
Shani \etal \cite{Shani16} considered a recommendation system based on
Markov decision processes (MDP). For this, they used the Maximum Likelihood Estimates
(MLE) of the MC transition graphs and they suggested several
heuristic approaches such as clustering and skipping.
Mobasher \etal \cite{Mobasher02} adopted pattern mining methods to extract
sequential patterns for generating recommendations.
He \etal \cite{He2018} proposed a Translation-based Recommendation (TransRec) with sequential data. The main approach is to consider items(movies) as a translation vector. Khorasani \etal \cite{Khorasani2016} also used the MC to recommend courses that students taken.
To do this, they estimated the transition probability of the MC from the record of courses students take based on MLE and enhanced MLE with skip-gram modeling.
Konen \etal \cite{Koren2010} considered temporal dynamics and
they showed several results by using
the evolution of users and items over time-based on Netflix data.

For the long-term dynamics, most recommendation usually relies
on Matrix Factorization (MF) or other similarity-based approaches.
In the prior work \cite{Koren2009} using the MF, the authors considered the recommendation
problem as a problem that infers missing values of
a partially observed user–item matrix. Srebro \etal \cite{Srebro2005} proposed the maximum margin
MF, which used low-norm instead of low-rank factorizations.
Salakhutdinov \etal \cite{Mnih2017} considered a probabilistic
MF (PMF) model that expresses the user preference matrix
by multiplication of two lower-rank user and item matrices.
The PMF approach was especially effective at making better
predictions for sparse user rating data.
He \etal \cite{He2016} suggested an extended FPMC, called Fossil, to present the information
of sequential patterns by considering high-order Markov
chains and similarity models. As
factorization machine-based sequential recommendation systems
usually utilize the matrix or tensor factorization to factorize the observed user-item related data into latent factors
of users and items for recommendations \cite{Hidasi2018}.
Specifically, some works \cite{Wang2015,Wang2018}
have used the estimated latent representations as an input of a network
to further calculate an interaction score between users
and items, or successive users’ actions.

Recently, deep learning technologies have been introduced in the sequential recommendation problem, such as RNNs \cite{Hidasi2018,Dong2018,Chairatanakul2019}, Long-Short Term Memory (LSTM) \cite{Wu2017,Zhao2020}, and Gated Recurrent Unit (GRU) \cite{Hidasi20161}. These deep learning-based recommender systems have particularly shown a high performance
for the sequential recommendation. In \cite{Hidasi2018}, the authors suggested new ranking loss functions corresponding to RNNs in the recommendation model. In \cite{Dong2018} the authors designed a novel recommendation model named Recurrent Collaborative Filtering (RCF), which combines RNN and CF properly. In \cite{Chairatanakul2019} the authors introduced an algorithm named Recurrent Translation-based Network
(RTN). Their model reflected both short-term and long-term of
a user’s preference. In \cite{Wu2017} the authors considered LSTM to extract the dependencies of both users and movies. Unlike the prior recommendation models, they considered a method of updating the state with recent operations as input. In \cite{Zhao2020} the authors introduced an LSIC model, Leveraging long and Short-term Information in
Content-aware movie recommendation via
adversarial training, which combined global
behaviors from MF into the RNN for the top-$N$ movies.
\label{sec:method}
\begin{figure*}[t!]
\begin{center} \centering
\includegraphics[width=1.1\linewidth]{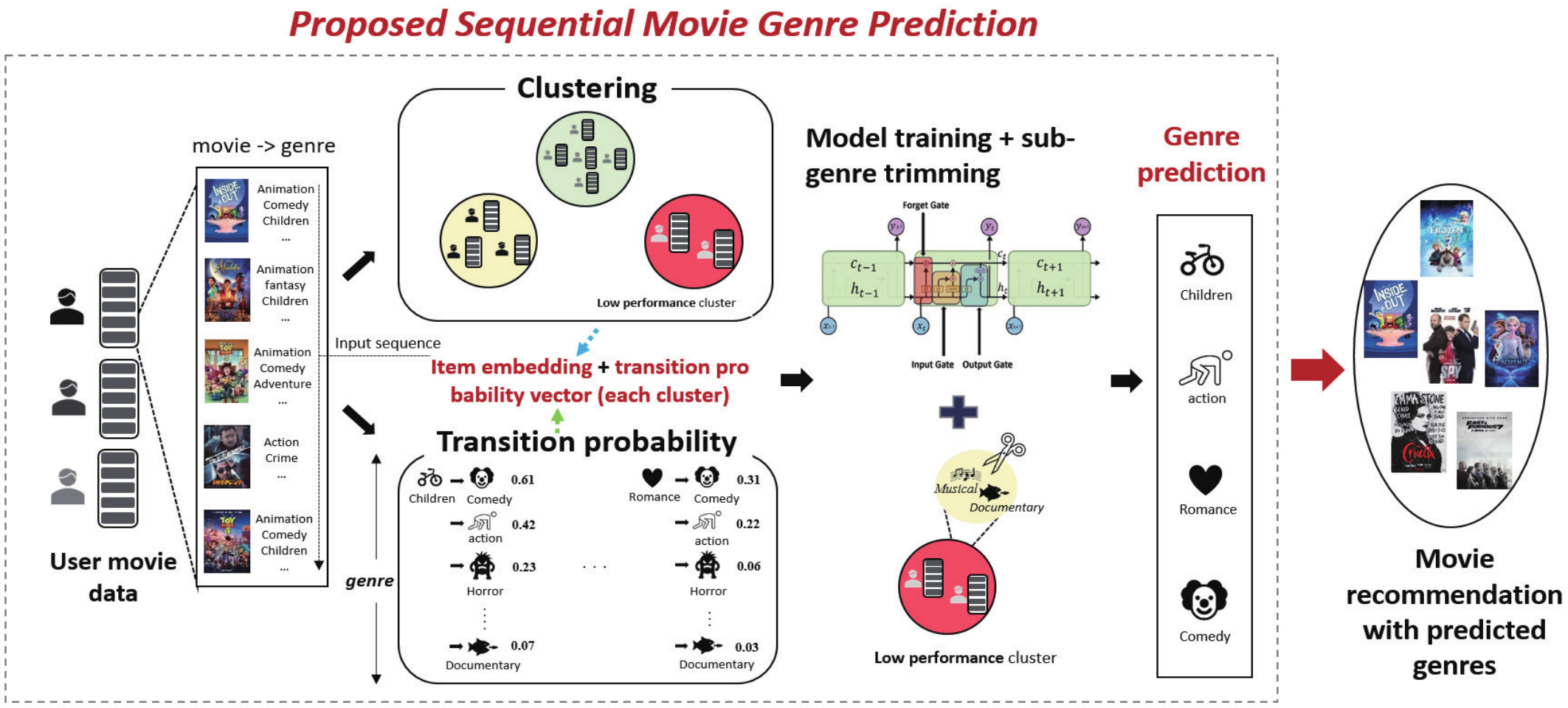}
\caption{Overall view of our proposed sequential movie genre prediction. }
\label{fig:overall}
\vspace{-0.6cm}
\end{center}
\end{figure*}
In \cite{Hidasi20161} the authors considered a GRU-based RNN for session-based recommendations.
Yuan \etal \cite{Yuan2019} suggested a Convolutional Neural Network (CNN) method that gives a sequence
of user-item interactions. In their model, a CNN first puts
user-item interactions data into a matrix, regarding such a matrix
as an “image” in the time and latent spaces. Wu \etal \cite{Wu2019} proposed a GNN
to capture the sequential behavior of complex transitions over user-item
interactions. Zhang \etal \cite{Zhang2019} adopted a self-attention
mechanism to extract the item-item interaction from the user’s historical interactions. Sachdeva \etal \cite{achdeva2019} considered a variational autoencoder to model a user’s preference
based on her historical sequential data, and combines latent variables with temporal dependencies
for preference modeling. Similar to our works, Choi \etal \cite{Choi2012} designed a movie recommendation algorithm based on genre correlations. For this, they assumed that movie genres are defined by some experts such as directors or producers to guarantee reliability. Then, they computed genre correlations and used them in a movie recommendation system. In \cite{Kim2012}, the authors also consider a movie genre similarity to provide related services in a mobile experimental environment.

\section{Genre prediction Algorithm}

In this section, we will propose a movie genre prediction algorithm. To do this, we first classify the genres included in the movie data watched by each user as shown in Figure~\ref{fig:overall}. Next, we cluster the users into similar groups based on the ratings of the movies. Then, we estimate an average transition probability from genre to the genre for each cluster. Using this, we train some deep learning models. Since some sparse data of genres may cause poor performance in predicting the genres, we appropriate trim these after model training, and we finally predict a preferred genre for the group closest to the user. Based on the predicted genre, some suitable movies that contain that genre can be recommended. We describe all the above steps in detail as the following subsections.

\subsection{Data Preprocessing}
\begin{figure}[H]
\begin{center} \centering
\includegraphics[width=1\linewidth]{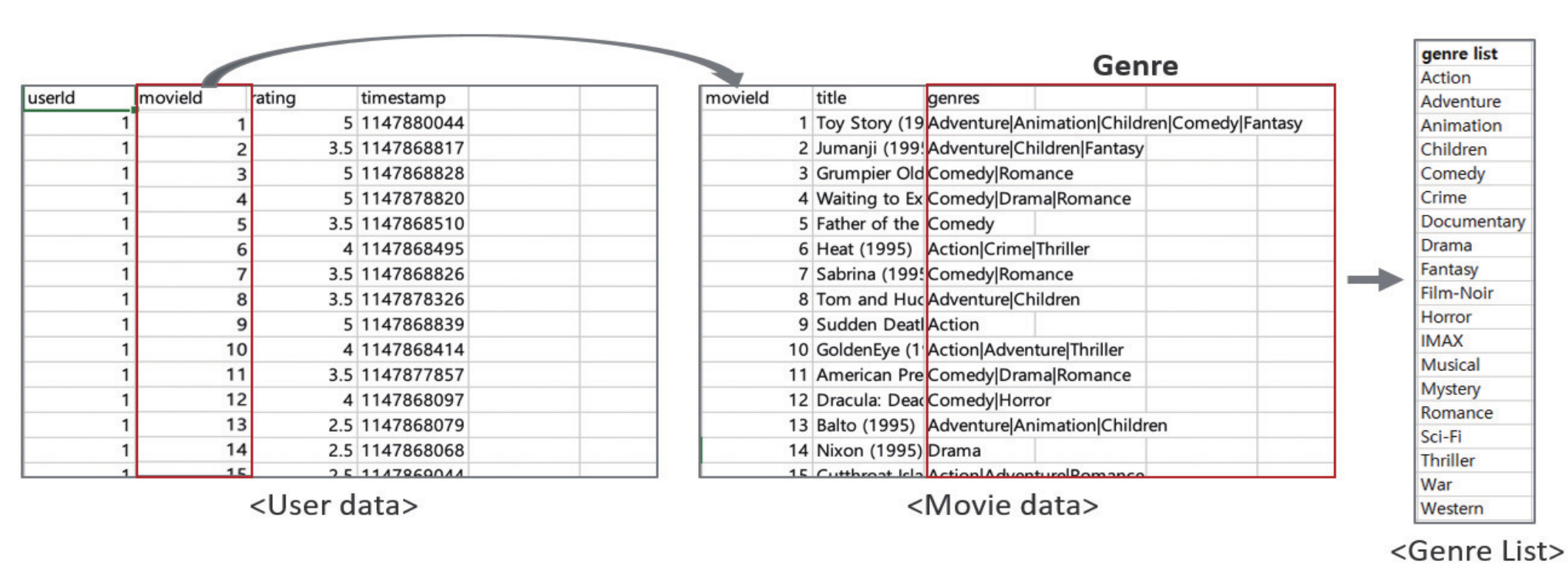}
\caption{Example of data preprocessing for movie genre prediction ($n=19$).}
\label{fig:data}
\vspace{-0.6cm}
\end{center}
\end{figure}

As a sequential movie genre prediction, we consider movie data by user ID and timestamp to extract each user's movie sequence in chronological order (left part of Figure~\ref{fig:data}). We drop user data with five or fewer movie viewing sequences and import user data with five or more movie viewing sequences in the pre-processing. At this time, the five most recent movies generated per user are arranged in chronological order. Next, information on genres included in movies that the user has watched is organized (middle part of Figure~\ref{fig:data}).
One can see that a single movie contains several genres at the same time. We extract all kinds of genres that each film contains and set the data sequence to $n$-dimensions ($n>0$) of one-hot vector, meaning each of the $n$ genres (right part of Figure~\ref{fig:data}). We denote $\mathcal{G}$ the set of genres in the paper.

\smallskip
\subsection{Clustering}
\begin{figure}[H]
\begin{center} \centering
\includegraphics[width=1\linewidth]{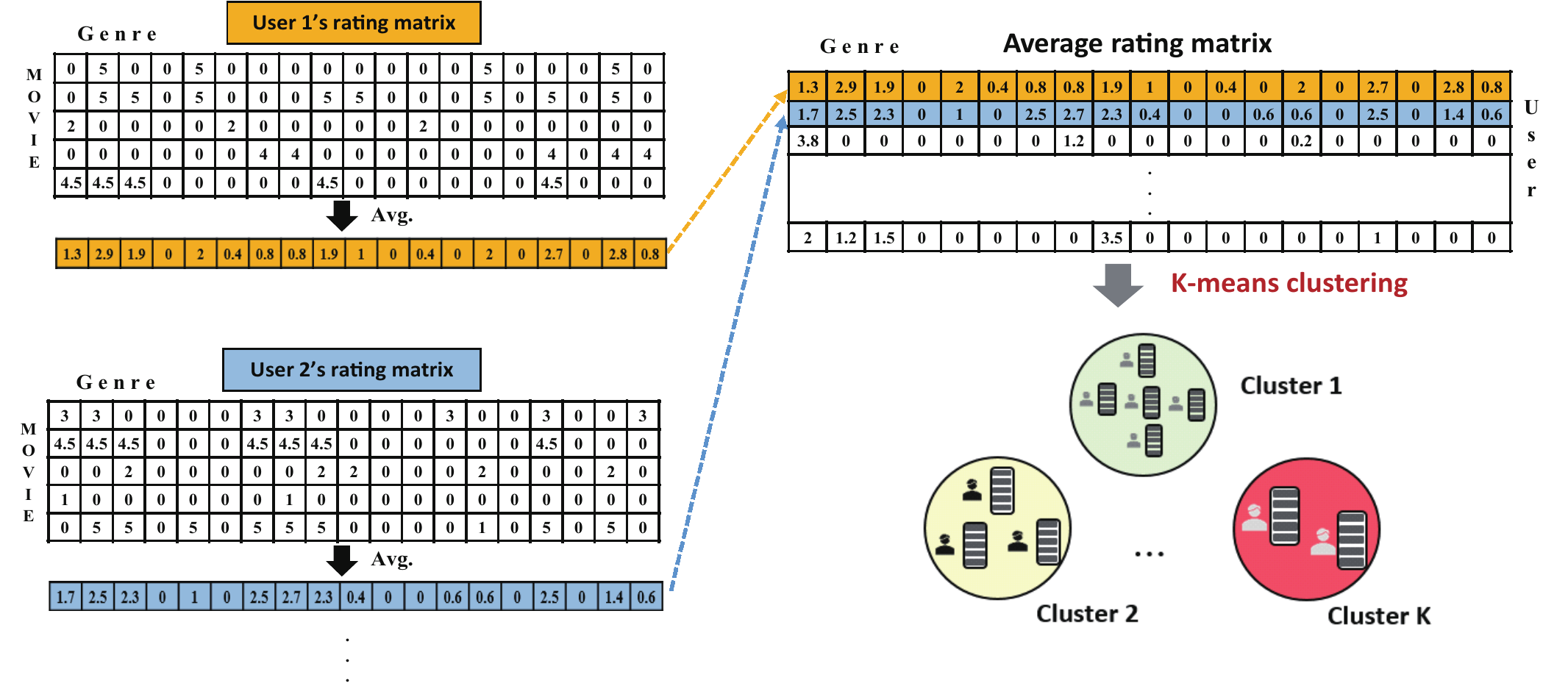}
\caption{Illustration of example for clustering.}
\label{fig:clustering0}
\vspace{-0.6cm}
\end{center}
\end{figure}

To reflect user similarity, we consider a clustering approach. For this, we consider that each user scores a range between 0.5 and 5 rating for the most recent five movies they watched. Based on the rating sequences of each user, we obtain the average rating of each genre as shown in Figure~\ref{fig:clustering0}. Let $\mathcal{U}$ be the set of users and we consider an average rating data is generated per one user so we have $|\mathcal{U}|$ by $n$ rating matrix. Using this matrix, we apply a $k$-means clustering to obtain clusters. We let $\mathcal{C}:=\{C_1,...,C_k\}$ be a set of clusters $C_l$ for $1 \leq l \leq k$ after performing the clustering.

\smallskip

\smallskip
\subsection{Average Transition Probability of genres}
\smallskip
\subsubsection{ Markov Chain for Set of Genres.}
\begin{figure}[H]
\begin{center} \centering
\includegraphics[width=1\linewidth]{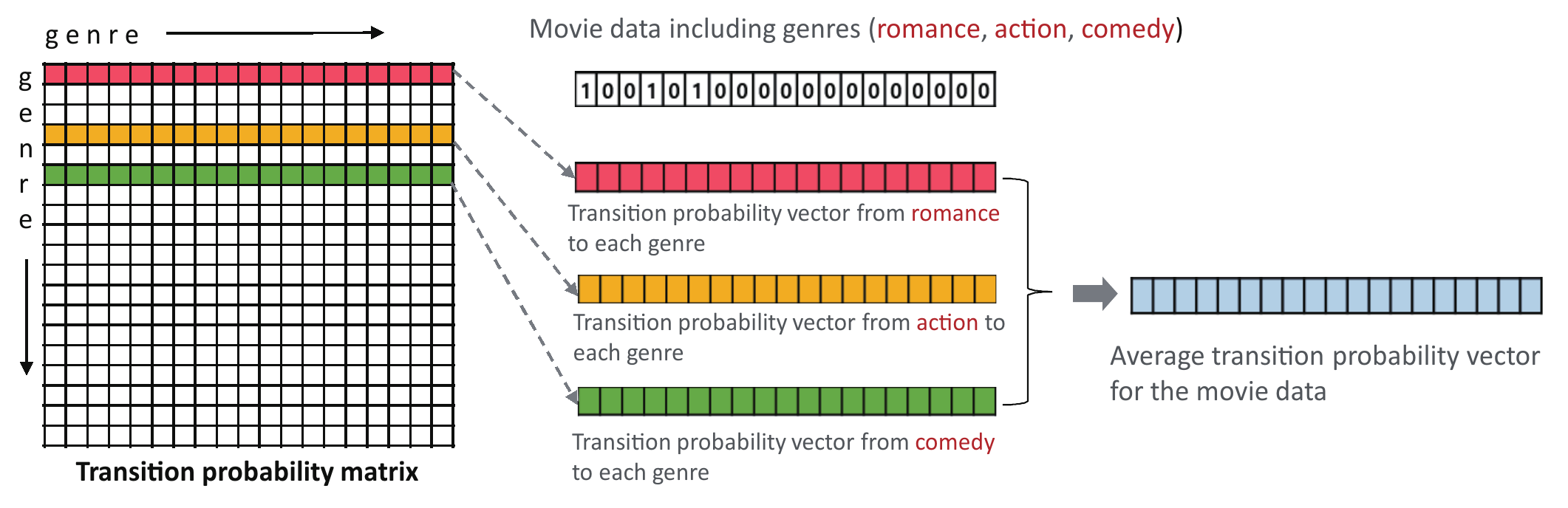}
\caption{Transition probability matrix and average transition probability vector.}
\label{fig:transition}
\vspace{-0.2cm}
\end{center}
\end{figure}

In our genre prediction system, we use transition probabilities from genre to genre. It is known that many approaches for the sequential recommendation, the MC is used to reflect the short-term sequential behavior of a user. The MC assumes the next choice of item depends only on the current choice. Formally, it is described as follows. The transition probability matrix is generated for each cluster. To do this, we first consider the sequence of selected movies for each user in a cluster. However, as described before, a movie may contain multiple genres such as romance, action and comedy, simultaneously. We consider all genres included in the current and next movies and count them in the $n$ by $n$ matrix. We consider the transition matrix for each cluster, separately. Then, for all $C_l \in \mathcal{C}$, we summarize them for all user's chosen movies and normalize them to obtain the transition probability from genre to genre as shown in Figure~\ref{fig:transition}.

\smallskip

To describe this, we let $M^{l}_{t}$ and $M^{l}_{t-1}$ be the selected movie sets for all user $u \in C_l$ at time $t$ and $t-1$, respectively. Then, the transition probability of the first-order Markov chain for the movie selection for the cluster $l$ is given by:
\begin{align}
      \label{eqn:probability}
      p(M^{l}_{t} | M^{l}_{t-1}).
 \end{align}
However, in the genre prediction, we focus on the transition from genre (included in a movie) to genre. For this, we let $\mathcal{G}_{t} \subset \mathcal{G}$ be the set of genres which are contained in the movie $M^{l}_{t}$ for all user $u \in C_l$ at time $t.$ Consider two genres $i,j \in \mathcal{G}$, we model the genre transition probability in the cluster $C_l$ as:
\begin{align}
      \label{eqn:transition}
      p_{ij}:= p(j \in \mathcal{G}^{l}_{t} | i \in \mathcal{G}^{l}_{t-1}).
 \end{align}

%\smallskip
\subsubsection{ Estimation of Transition Probabilities.}
To make predictions using the transition probability in \eqref{eqn:transition}, it needs to be estimated. To do this, we consider the following ratio:
\begin{align}
      \label{eqn:transition1}
      \hat{p}_{ij}:= \hat{p}(j \in \mathcal{G}^{l}_{t} | i \in \mathcal{G}^{l}_{t-1})&=\frac{\hat{p}(j \in \mathcal{G}^{l}_{t} \wedge i \in \mathcal{G}^{l}_{t-1})}{\hat{p}(i \in \mathcal{G}^{l}_{t-1})}\\
      &= \frac{|\{(\mathcal{G}^{l}_{t},\mathcal{G}^{l}_{t-1}): j \in \mathcal{G}^{l}_{t} \wedge i \in \mathcal{G}^{l}_{t-1}\}|}{|\{(\mathcal{G}^{l}_{t},\mathcal{G}^{l}_{t-1}): i \in \mathcal{G}^{l}_{t-1}\}|},
      \label{eqn:transition2}
 \end{align}
where the value of the denominator in \eqref{eqn:transition2} is the number of genre $i$ at time $t-1,$ and the numerator means the number of genre $i$ at time $t-1$ and genre $j$ at the next time point $t$. Hence, the estimated transition probability indicates that the ratio that the number of genre $j$, which is selected at time $t$ among number of genres $i$ at time $t-1.$ However, since the user does not select a specific genre, but sequentially selects a movie including several genres, the transition probability between genres cannot be used by itself.
Hence, using the movie data including several genres, we count the number of transitions to each genre. For example, if a movie contains three genres (romance, action and comedy) as in Figure~\ref{fig:transition}, count all genres which are included in the next selected movie. Then we compute the ratio in \eqref{eqn:transition2} so that we have the estimated transition probabilities from romance to each genre, from action to each genre and from comedy to each genre, respectively. Next, we take an average for these tree transition probabilities and call it an Average Transition probability Vector (ATV) for each selected movie. The reason why we use the ATV is that there is no information about transition from a specific genre to another genre in actual data, only information about transition from a movie including these genres to another movie is given. Formally, the ATV can be presented by:
\begin{align}
\hat{p}_{j}^{ATV}:=p(j \in \mathcal{G}^{l}_{t}| \mathcal{G}^{l}_{t-1})=\frac{1}{|\mathcal{G}^{l}_{t-1}|}\sum_{i \in \mathcal{G}^{l}_{t-1}}p(j \in \mathcal{G}^{l}_{t}| i \in \mathcal{G}^{l}_{t-1}),
 \end{align}
for all $j \in \mathcal{G}.$ We will use this ATV for training with each user's selected movie sequence.

\subsection{Model Training}
\subsubsection{Training Data Types}
As training data, we consider the following four types of training data during the model training: (1) Sum of transition vector and movie genre embedding, (2) Multiplication of transition vector and movie genre embedding (3) Successive transition vector and movie genre embedding, and (4) Movie genre only.
\begin{figure}[H]
\begin{center} \centering
\includegraphics[width=1\linewidth]{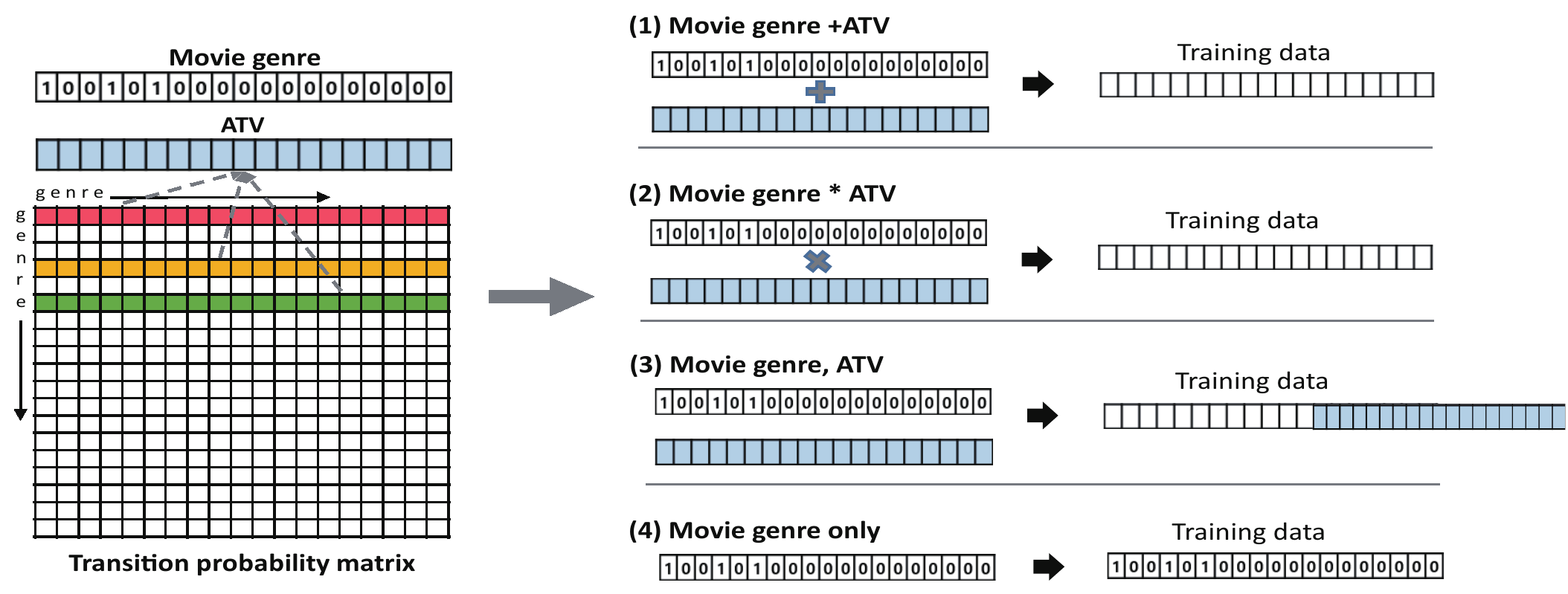}
\caption{Four types of training data}
\label{fig:datatypes}
\vspace{-0.2cm}
\end{center}
\end{figure}

First, the sum of ATV and movie genre embedding data is nothing but performing the summation of movie genre vector and ATV as shown in Figure~\ref{fig:datatypes}. Second, the multiplication of ATV and movie genre embedding is the data after multiplying these two vectors by component-wise, which results in a new vector. Third, the successive ATV and movie genre embedding is the data that attaches the ATV at the end of the movie genre for the model training. Finally, the movie genre only is the data that consider only the movie genre vector. The reason we consider the training data types in this way is to check how much the ATV considered for short-term dynamics helps to improve the model performance. We will show the results for these four training data types in the experiment later.

\subsubsection{Training Models} In our approach, we use RNN-based models to capture the long-term dynamics of sequential movie genre data such as RNN, LSTM and GRU. We will describe these methods in detail as follows.

\smallskip
{\bf \em (1) RNN.} First, RNN is one of the deep learning models designed to be useful for sequential data processing. RNN is a recursive model that performs the same function on all input data and the output for the current input depends on past calculations. When the output data is generated, it is copied and sent back into the recurrent network.
Based on the current input and the output that it has generated from the previous input,
the RNN learns some sequential data and makes a decision.
\begin{figure}[H]
\begin{center} \centering
\includegraphics[width=0.8\linewidth]{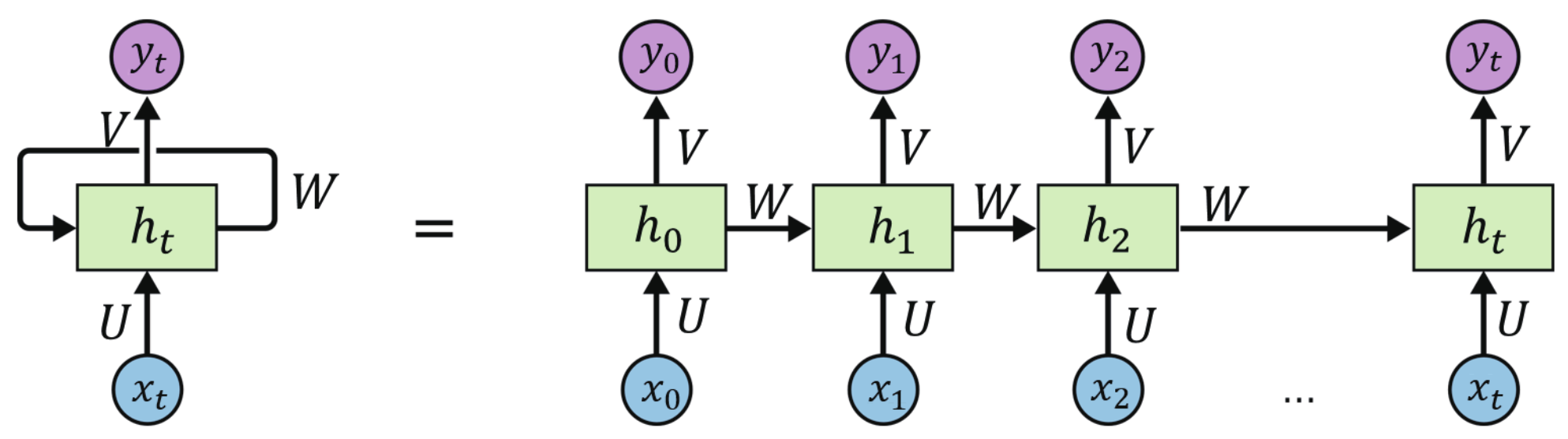}
\caption{Recurrent Neural Network.}
\label{figs:RNN}
\vspace{-0.2cm}
\end{center}
\end{figure}

% \smallskip
% \begin{figure}[H]
% \begin{center}
% \subfigure[~Recurrent Neural Network.]{\includegraphics[width=0.8\columnwidth]{figs/RNN.eps}\label{figs:RNN}}
% \hspace{-0.25cm}
% \subfigure[~
% Long Short Term Memory.]{\includegraphics[width=0.79\columnwidth]{figs/LSTM.eps}\label{fig:LSTM}}
% \end{center}
% \caption{RNN and LSTM.}
% \label{fig:BSRA}
% \vspace{-0.2cm}
% \end{figure}
% \begin{figure}[H]
% \begin{center} \centering
% \includegraphics[width=0.78\linewidth]{figs/LSTM}
% \caption{LSTM as the learning model.}
% \label{fig:ER}
% \vspace{-0.2cm}
% \end{center}
% \end{figure}
To formally describe, we let $x_t$ be the input vector and $y_t$ be the output vector at time $t$ as shown in Figure~\ref{figs:RNN}. Then, a state value of hidden layer $h_t$ at time $t$ is given by:
\begin{align}
      \label{eqn:hidden}
     h_t = \tanh(Ux_t + Wh_{t-1} + b),
 \end{align}
where $U, W$ are model parameter matrix and $b$ is a constant vector. As a function of $h_t$, we consider a hypublic tangential function $tanh(\cdot)$. The output vector $y_t$ is given by:
\begin{align}
      \label{eqn:hidden}
     y_t = f(Vh_{t} + b),
 \end{align}
where $V$ is a model parameter and $f$ is an activation function. RNN is optimized to approximate the function by capturing
sequential patterns. However,
if the length of the sequence input to the RNN is long, the effect of the elements at the beginning of the sequence will gradually loosing as the time step progresses and disappear after a certain period of time. This is because the constant value is multiplied equally in each cycle.
This is called a long-term dependency problem that the RNN is useful for a short sequence of data.

\smallskip
{\bf \em (2) LSTM.}
To overcome the main disadvantage of RNN, one of the improved methods, LSTM has been introduced. LSTM \cite{LSTM} is a kind of RNN that is capable of selectively remembering sequences for a long period of time. The main difference from the RNN is that LSTM introduces a ``cell state" for each time $t$, which allows information to flow unaltered. In LSTM, the cell state is regarded as a long-term memory since the previous information is stored in it as a recursive nature of the cells. The forget gate is used to update the cell states. The forget gate outputs values saying which information to forget by multiplying 0 to a position in the matrix. If the output of the forget gate is 1, the information is kept in the cell. The input gates determine which information should enter the cell states. Finally, the output gate tells which information should be passed on to the next hidden state. Based on this fact, the LSTM addresses the long-term dependency problem of RNN.
In general, the LSTM consists of the following four parts as shown in Figure~\ref{fig:LSTM}:
\begin{itemize}
\item[$(i)$]
%\smallskip
{\bf \em Forget Gate Layer.}
As a first part, the forget gate layer decides to filter some information from the cell state by using a sigmoid function. It obtains information at $h_{t-1}$ and $x_t$, and outputs a number between 0 and 1 for each number in the cell state $c_{t-1}$. The number 1 implies “completely keep this” while 0 represents “completely drop this.” The output of the forget gate vector $t_t$ is given by:
\begin{align}
      \label{eqn:forget}
     f_{t}=\sigma(W_f \cdot [h_{t-1},x_t]+b_f),
 \end{align}
where $\sigma$ is a sigmoid function and $W_f$ and $b_f$ are weight matrix and bias vector parameter.
\begin{figure}[H]
\begin{center} \centering
\includegraphics[width=0.8\linewidth]{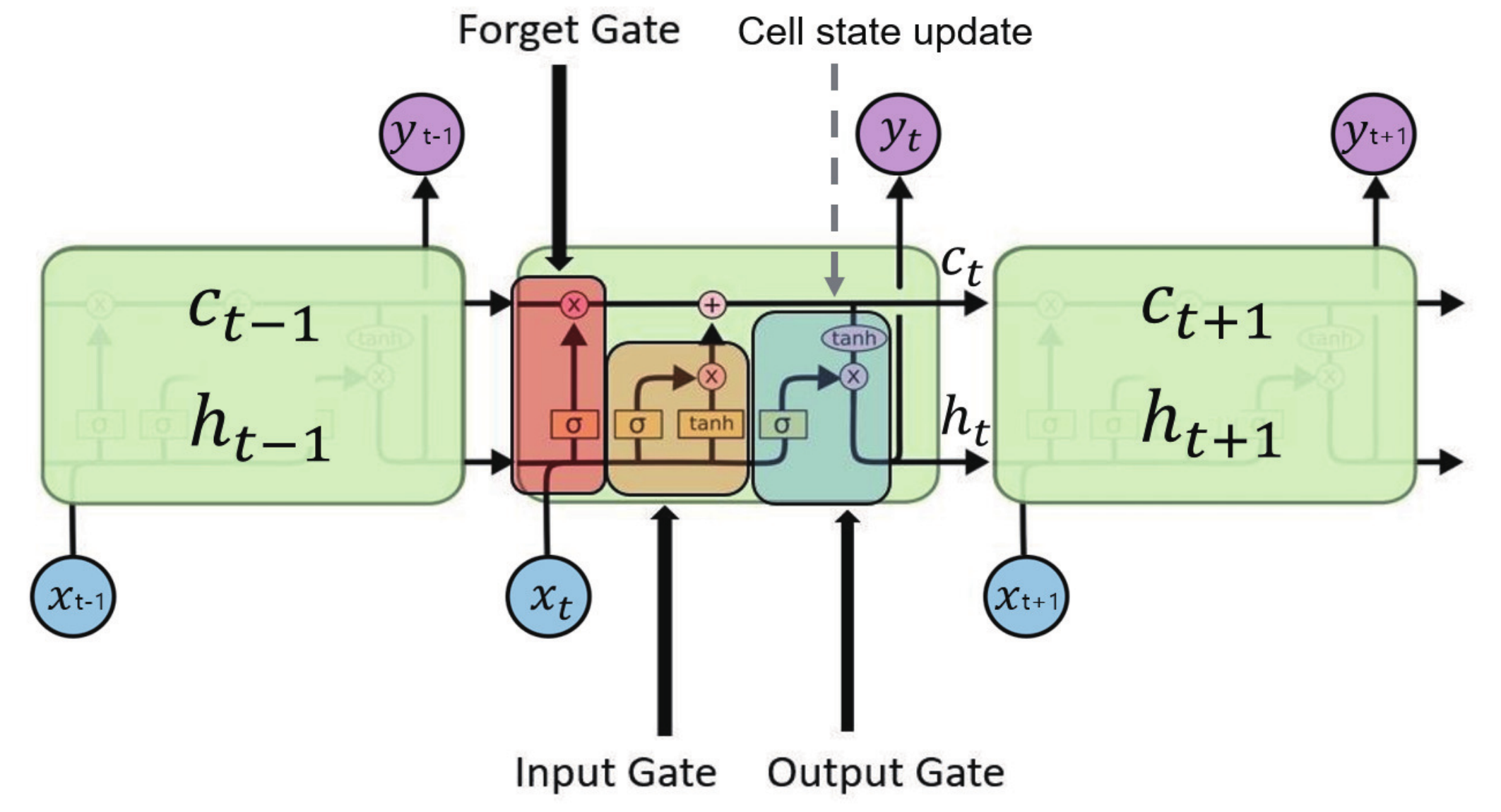}
\caption{Long Short Term Memory \cite{LSTM}.}
\label{fig:LSTM}
\vspace{-0.2cm}
\end{center}
\end{figure}
\item[$(ii)$]
{\bf \em Input Gate Layer.} In the next step, LSTM decides whether new information to store or not in the cell state. For this, an “input gate layer” decides which values we’ll update as a sigmoid gate.
Next, a tanh gate generates a vector of new values, $\tilde{c}_t$, that could be added to the state. Then, these two layers are combined to create an update to the state. The input gate vector $i_t$ is given by:
\begin{align}
      \label{eqn:input1}
     i_{t}=\sigma(W_i \cdot [h_{t-1},x_t]+b_i),
 \end{align}
where $W_i$ and $b_i$ are weight matrix and bias vector parameter. The cell input activation vector $\tilde{c}_t$ is computed by:
\begin{align}
      \label{eqn:input2}
     \tilde{c}_t =\tanh(W_c \cdot [h_{t-1},x_t]+b_c),
 \end{align}
 where $W_c$ and $b_c$ are weight matrix and bias vector parameter and $\tanh(\cdot)$ is a hyperbolic tangential function as a sigmiod function.

  %\smallskip
\item[$(iii)$]
{\bf \em Cell State Update.}
Next, LSTM performs a cell state update procedure to update the old cell state, $c_{t-1}$, into the new cell state $c_{t}$. The previous steps already decided what to do, it just needs to actually do it.
Then, it multiplies the old state to $f_{t}$, forgetting the things it decided to forget earlier. Then it adds $i_{t} \cdot \tilde{c}_t$. This is the new candidate value, scaled by how much it decided to update each state value. The update of cell state $_t$ is computed by:
\begin{align}
      \label{eqn:input2}
     c_t = f_t \ast c_{t-1} + i_{t} \ast \tilde{c}_t.
 \end{align}

  %\smallskip
\item[$(iv)$]
{\bf \em Output Gate Layer.}
Finally, in the output gate layer, LSTM decides what information going to be output. This output will be based on the cell state, but will be a filtered version. First, it runs a sigmoid layer which decides what parts of the cell state going to output. Then, it puts the cell state through tanh and multiplies it by the output of the sigmoid gate, so that it only output the parts it decided to. The output gate vector $o_{t}$ is given by:
\begin{align}
      \label{eqn:output1}
     o_{t}=\sigma(W_o \cdot [h_{t-1},x_t]+b_o),
 \end{align}
where $W_o$ and $b_o$ are weight matrix and bias vector parameter of the output gate layer. Here, $h_t$ is computed by:
\begin{align}
      \label{eqn:output2}
     h_t =o_t \ast \tanh(c_t).
 \end{align}

\end{itemize}

\smallskip
{\bf \em (3) GRU.}
Cho \etal \cite{Cho2014} first introduced a slight variation on the LSTM, named GRU. It uses the forget and input gates as a single update gate. Further, it also combines the cell state and hidden state. It is known that the GRU is simpler than LSTM model.
\begin{figure}[H]
\begin{center} \centering
\includegraphics[width=0.7\linewidth]{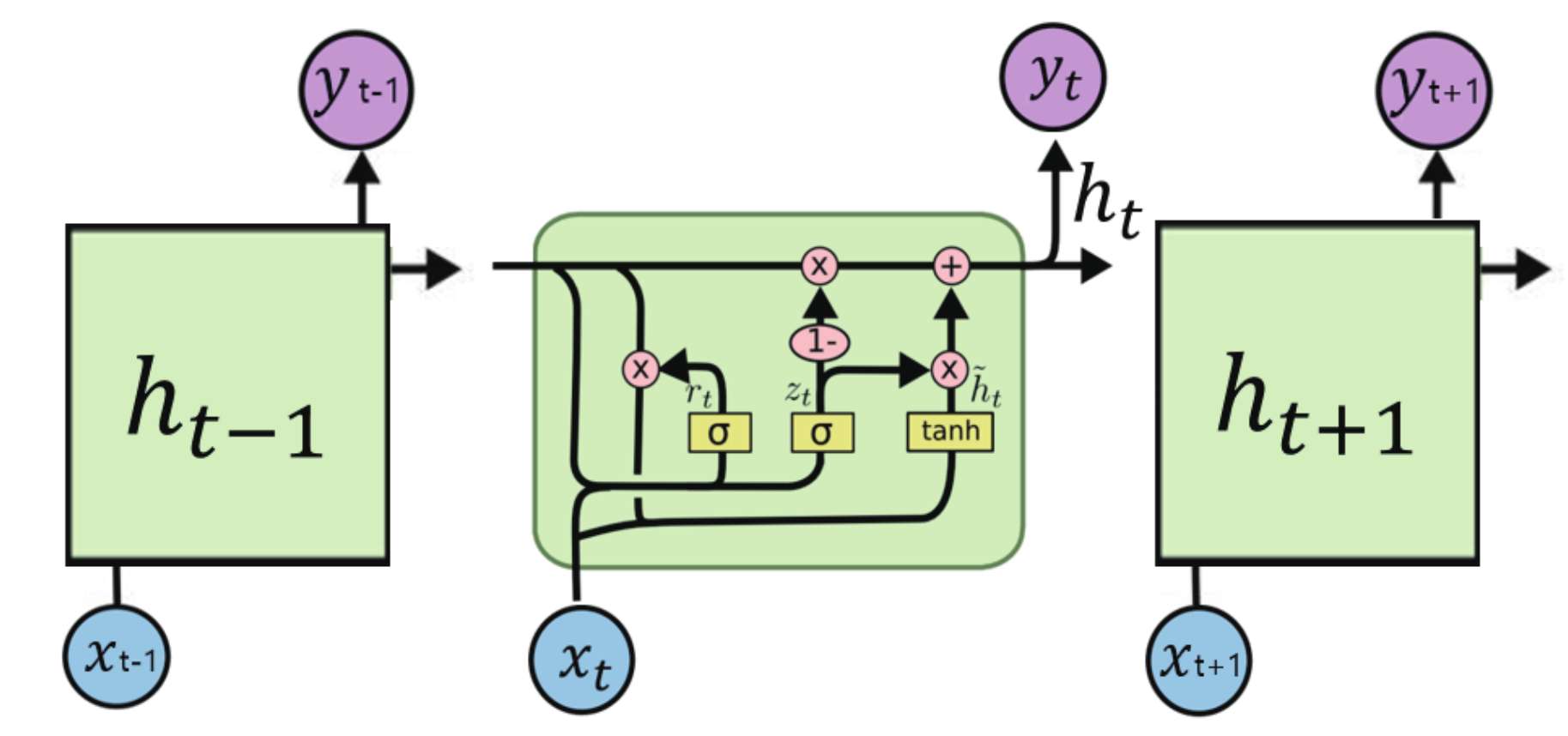}
\caption{GRU~\cite{GRU}.}
\label{fig:GRU}
\vspace{-0.2cm}
\end{center}
\end{figure}
\smallskip

The detailed structure of GRU as shown in Figure~\ref{fig:GRU} is in what follows:
\begin{itemize}
\item[$(i)$] {\bf \em Update Gate.} In GRU \cite{GRU}, it first begins with computing the update gate $z_t$ for time step $t$ by:
\begin{align}
      \label{eqn:out1}
     z_{t}=\sigma(W_z \cdot [h_{t-1},x_t]),
 \end{align}
where $W_z$ is a weight matrix. When the input $x_t$ is generated into the network, it is multiplied by its own weight $W_z$. The previous $h_{t-1}$ also multiplied by the current input $x_t$. As an activation function, a sigmoid is commonly used. The update gate is used to determine how much of the past information (from previous time steps) needs to be passed along to the next. The most useful fact is that the model can control to copy all the information from the past and eliminate the risk of vanishing gradient problems.

 \item[$(ii)$] {\bf \em Reset Gate.} Next, a reset gate is applied from the model to decide how much of the past information to forget by:
 \begin{align}
      \label{eqn:out2}
     r_{t}=\sigma(W_r \cdot [h_{t-1},x_t]),
 \end{align}
The difference between this from the update gate is the weights and the gate’s usage. As similar steps in the update gate, it plugs in $h_{t-1}$ and $x_t$, multiply them with their corresponding weights, sum the results and apply the sigmoid function.

 \item[$(iii)$] {\bf \em Current memory content.} The current memory content is then used for the reset gate to store the relevant information from the past. It is computed as:
  \begin{align}
      \label{eqn:out3}
     \tilde{h}_{t}=\tanh(W \cdot [r_t \ast h_{t-1},x_t]),
 \end{align}
 where $W$ is a weight matrix and the operator $*$ denotes the Hadamard element-wise product. Then the result determines what to remove from the previous time steps.
In this step, it uses a tanh as the nonlinear activation function.

\item[$(iv)$] {\bf \em Final memory at current time step.} As the last step, the network needs to calculate $h_t$, which is a vector that holds information for the current unit and passes it down to the network. In order to do that the update gate is needed. It determines what to collect from the current memory content $\tilde{h}_{t}$ and what from the previous steps $h_{t-1}$ by weighting the update gate value:
 \begin{align}
      \label{eqn:out4}
     h_t =(1-z_t) \ast h_{t-1} + z_t \ast \tilde{h}_{t}.
 \end{align}
 \end{itemize}

\subsubsection{Sub-genre trimming}
Finally, using the evaluation results of the trained models, we perform a sub-genre trimming process based on a pre-defined threshold of the evaluation metric scores for each cluster. To do this, we first select clusters that do not satisfy the criteria of evaluation metrics.
\begin{figure}[H]
\begin{center} \centering
\includegraphics[width=1\linewidth]{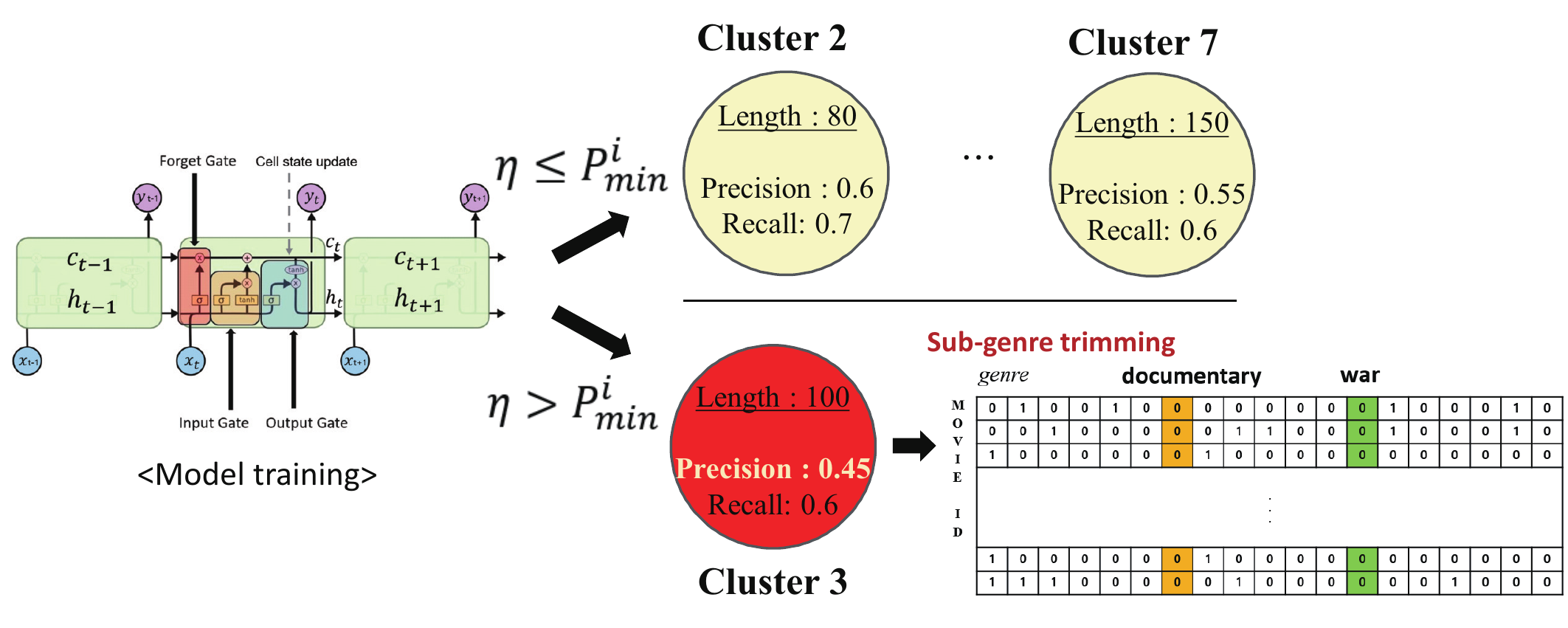}
\caption{Sub-genre trimming. In this example, we set $\eta$=0.5 and there are two kinds of performance metrics such as precision and recall. We see that the minimum value of evaluation metrics $P^{2}_{min} =0.6$ and $P^{7}_{min} =0.55$ for the cluster 2 and cluster 7, respectively. Hence, these clusters are not regarded as the trimming clusters, However, we have $P^{3}_{min} =0.45<0.5$ for cluster 3 so it is regarded as a trimming cluster.}
\label{fig:trimming0}
\vspace{-0.2cm}
\end{center}
\end{figure}

\begin{algorithm}[H]
\caption{{Sub-genre Trimming}}\label{alg:NB}
\begin{algorithmic}
   \State \textbf{Input:}  Set of movie-genre matrices $\mathcal{M}:=\{\mathcal{M}_i\}^{k}_{i=1}$ for each cluster $C_i$ with each evaluated value $P^{i}_{e}$, threshold parameters $\eta$ and $\theta.$
  \State \textbf{Output:}  Sub-genre trimmed matrices set $\mathcal{M}^{'}:=\{\mathcal{M}_1^{'},...,\mathcal{M}_k^{'}\}.$

\smallskip

\State Set $C_i^{'}=\emptyset$ for all $1 \leq i \leq k$;

   \For {$1 \leq i \leq k$}
   \State Set $|C_i|$ the length (number of movies) of a cluster $C_i$ for each $i$ and set;
   \begin{align}
      \label{Precision}
    P^{i}_{min}:=\min_{e \in \mathcal{E}}P^{i}_e.
 \end{align}
   \If{$P_{min} < \eta$}
   \For {$1 \leq j \leq n$}
       \State Count the number of genre $j$ in a column of movie-genre matrix $\mathcal{M}_i$ and set it by $g_j = \sum_{u\in C_i} c_{uj}$. If $g_j <100 \times \theta_i$, replace all values of the column $j$ by zero;
     \EndFor\label{}
     \EndIf
    \State $\mathcal{M}^{'}_i\leftarrow \mathcal{M}_i$;
\EndFor\label{}
   \State  Return $\mathcal{M}^{'}:=\{\mathcal{M}_1^{'},...,\mathcal{M}_k^{'}\}$;
\end{algorithmic}
\vspace{-0.1cm}
\end{algorithm}

More precisely, we let $P^{i}_{e}$ be the value of evaluation for a performance metric $e \in \mathcal{E}$ of the cluster $i$, where $\mathcal{E}$ is a set of performance metrics such as $\mathcal{E}=\{Precision, Recall, Accuracy\}$. Next, we let $P^{i}_{min}:=\min_{e \in \mathcal{E}}P^{i}_e$ be the minimum evaluated value of $P_e$ for all $e \in \mathcal{E}$. Then, we check the value $P^{i}_{min} $ for each cluster and if there exists an evaluation value less than a pre-defined threshold $\eta>0$, \ie $P^{i}_{min} < \eta$, then we choose the cluster as the target cluster for the sub-genre trimming. For example, if we consider the evaluation metrics as precision and recall and the threshold $\eta$ is given by 0.5, the cluster 3 does not satisfies this as shown in Figure~\ref{fig:trimming0}. Hence, it is regarded as a target cluster for the sub-genre trimming.
 After selecting target clusters, we find sub-genres that are less than $\theta_i$ percent of the total length (number of movies) of each target cluster $i$ to trim. To do this, we let $\mathcal{M}_i=[c_{uj}]_{u \in C_i, 0 \leq j \leq n}$ be the movie-genre matrix for the cluster $i$ and let $\mathcal{M}:=\{\mathcal{M}_i\}^{k}_{i=1}$. Then, using the matrix, we find the genres that the number of total sum is less than $100 \times \theta_i$, \ie $\sum_{u\in C_i} c_{uj} < 100 \times \theta_i$. To minimize the data loss, we replace the values as zero rather than deleting thm. The reason for performing this is that it will increase the accuracy of evaluation of clusters that do not have explicit preference. In the example of Figure~\ref{fig:trimming}, the length of the cluster is 100 and $\theta=0.1$, then we have $100 \times 0.1 = 10$. We choose genres that do not have more than 10 data in the cluster such as documentary and war. After these procedures, we finally obtain the sub-genre trimmed matrices set $\mathcal{M}^{'}:=\{\mathcal{M}_1^{'},...,\mathcal{M}_k^{'}\}$. We will use this matrices to obtain the performance results.

\section{Experiment Results}\label{sec:simulation}
In this section, we will show our experiment results. For this, we first use a well-known movie dataset and performance metrics of the evaluation as follows.

\subsection{Data}
In this section, we present our experimential results. For the simulation, we use a movielens data set(ml-25m), where 25 million ratings and one million tag applications applied to 62,000 movies by 162,000 users \cite{movie}. For the sequential recommendation, we sort the data by ‘userId’ and ‘timestamp’ (to extract each user's movie sequence in chronological order) as shown in Figure~\ref{fig:data}. We drop user data with 5 or fewer movie viewing sequences and import user data with 5 or more movie viewing sequences to configure the dataset for the experiment. At this time, five movie data generated per user are arranged in chronological order. To train the model, we convert the data sequence to 19 dimensions of a one-hot vector, meaning each of the 19 genres: $\{Action, Adventure, Animation, Children, Comedy, Crime, Documentary, Drama, Fantasy,\\ Film-Noir, Horror, IMAX, Musical, Mystery, Romance, Sci-Fi, Thriller, War, Western\}.$

\subsection{Performance Metrics}
As performance metrics, we consider $(i)$ Recall, $(ii)$ Precision, $(iii)$ Accuracy and $(iv)$ F1-score. To formally explain these metrics, we denote True Positive (TP) as the number of correctly predicted positive values which is the actual value is yes and the predicted value is also yes. True Negatives (TN) indicates the number of correctly predicted negative values which is the actual value is no and the predicted value is also no. False Positives (FP) is the number of actual value is no and the predicted value is yes. False Negatives (FN) is the number of the actual value is yes but the predicted value is no. Then, the three metrics are described as follow:
\begin{itemize}
\item[$(i)$] {\bf \em Precision:} A Precision is the ratio of correctly predicted positive answers to the total predicted positive answers.
\begin{align}
      \label{Precision}
     Precision:=\frac{TP}{TP+FP}.
 \end{align}

\item[$(ii)$] {\bf \em Recall:} A Recall is the ratio of correctly predicted positive answers to all answers in the actual class of answers.
\begin{align}
      \label{Recall}
    Recall:= \frac{TP}{TP+FN}.
 \end{align}

\item[$(iii)$] {\bf \em Accuracy:} An Accuracy is a ratio of correctly predicted answers to the total answers.
\begin{align}
      \label{Accuracy}
     Accuracy:=\frac{TP+TN}{TP+FP+FN+TN}.
 \end{align}
The accuracy is one of good measures when the values of false positive and false negatives of the datasets are almost same.

\item[$(iv)$] {\bf \em F1-Score:} This metric is a weighted average of Precision and Recall. Therefore, this score takes both false positives and false negatives into account.
\begin{align}
      \label{F1}
    F1\text{-}score:= \frac{ 2(Recall \ast Precision)}{Recall + Precision}.
 \end{align}
F1-score is usually more useful than accuracy when the values of false positive and false negatives of the datasets are quite different.
\end{itemize}

Using the previously described data and performance metrics, we obtain various experimential results of the movie genre prediction in the following subsection.

\subsection{Results}
In the result, we obtain that how much the prediction performances are affected for (1) Clustering, (2) Sub-genre trimming, and (3) ATV. To see the clustering effect, we obtain the results as before and after clustering. We consider seven clusters after applying the $k$NN during the clustering step and the results show the mean performance of all clusters and best and worst performance of clusters among them, respectively. To select the trimming clusters, we set $\eta = 0.5$ and $\theta_i = 0.1$ for all cluster $i$.
\subsubsection{Effects of Clustering}

\begin{table}[H]
\caption{Four performance results of RNN with the training data type [Genre*ATV] }
\label{tab:RNN}
\centering
 \begin{tabular}{c|cccc}
\hline  Clustering    & Recall & Precision & Accuracy & F1-score \\
\hline BC &  0.44 %Is the bold necessary?
 & 0.77&0.85&0.56\\
\hline AC (best)& 0.73 & 0.80&0.91&0.76\\
\hline AC (worst)& 0.29 & 0.87&0.85&0.43\\
\hline AC (mean)& 0.52 & 0.77 & 0.86&0.58\\
\hline
\end{tabular} \end{table}

\begin{table}[H]
\caption{Four performance results of LSTM with the training data type [Genre*ATV] }
\label{tab:LSTM}
\centering
 \begin{tabular}{c|cccc}
\hline Clustering    & Recall & Precision & Accuracy & F1-score \\
\hline BC &  0.45 %Is the bold necessary?
 & 0.76&0.85&0.57\\
\hline AC (best)& 0.75 & 0.79&0.91&0.77\\
\hline AC (worst)& 0.27 & 0.84&0.84&0.41\\
\hline AC (mean)& 0.51 & 0.75 & 0.86&0.59\\
\hline
\end{tabular} \end{table}

\begin{table}[H]
\caption{Four performance results of GRU with the training data type [Genre*ATV] }
\label{tab:GRU}
\centering
 \begin{tabular}{c|cccc}
\hline  Clustering    & Recall & Precision & Accuracy & F1-score \\
 \hline BC &  0.49 %Is the bold necessary?
 & 0.70&0.83&0.58\\
\hline AC (best)& 0.73 & 0.75&0.89&0.74\\
\hline AC (worst)& 0.35 & 0.69&0.79&0.46\\
\hline AC (mean)& 0.53 & 0.69 & 0.84&0.59\\
\hline
\end{tabular} \end{table}

\begin{figure}[H]
\centering
\subfigure[~(RNN) Result without clustering.]{\includegraphics[width=0.4\columnwidth]{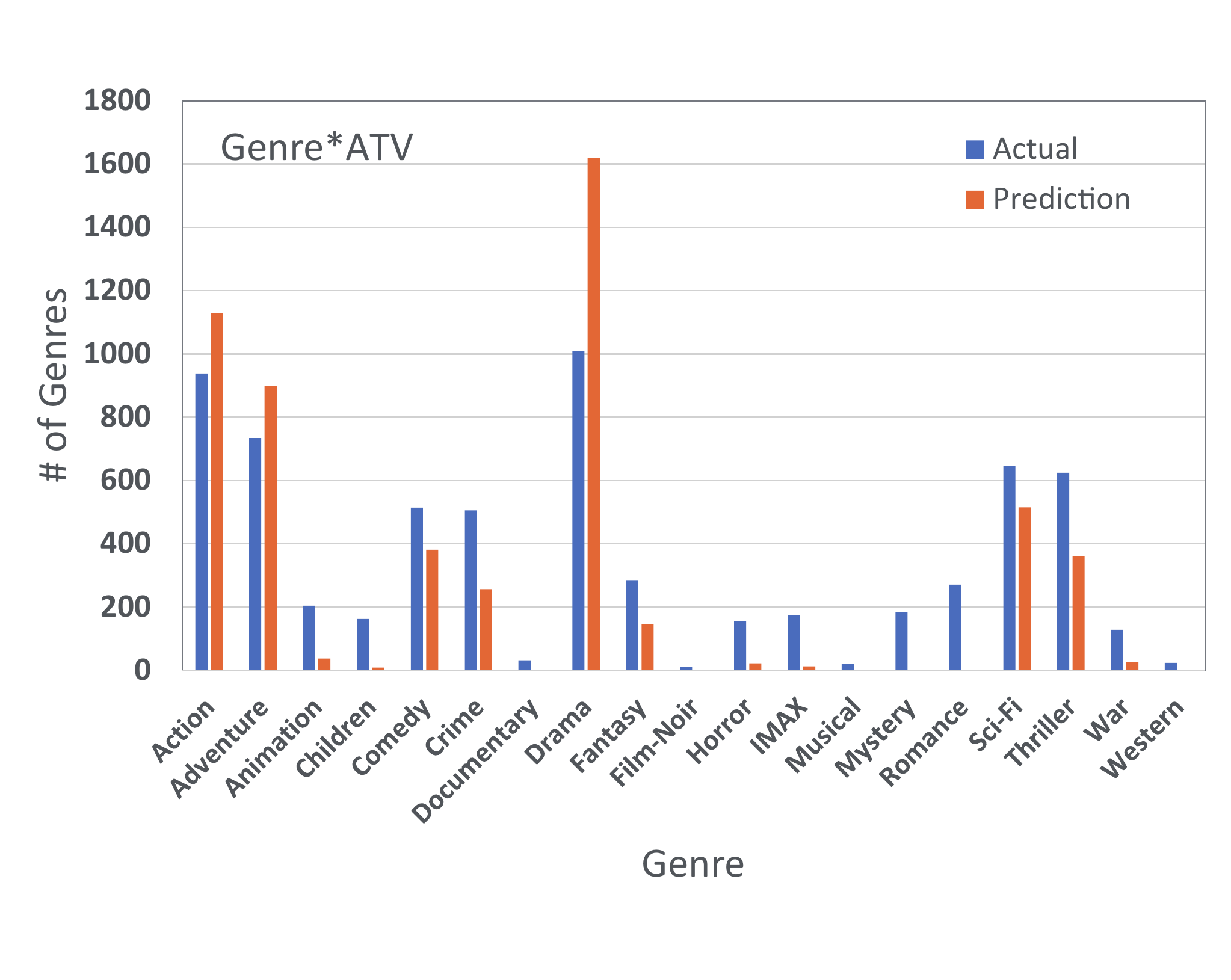}\label{fig:RNN1}}
\subfigure[~(RNN) Best result among all clusters after clustering.]{\includegraphics[width=0.4\columnwidth]{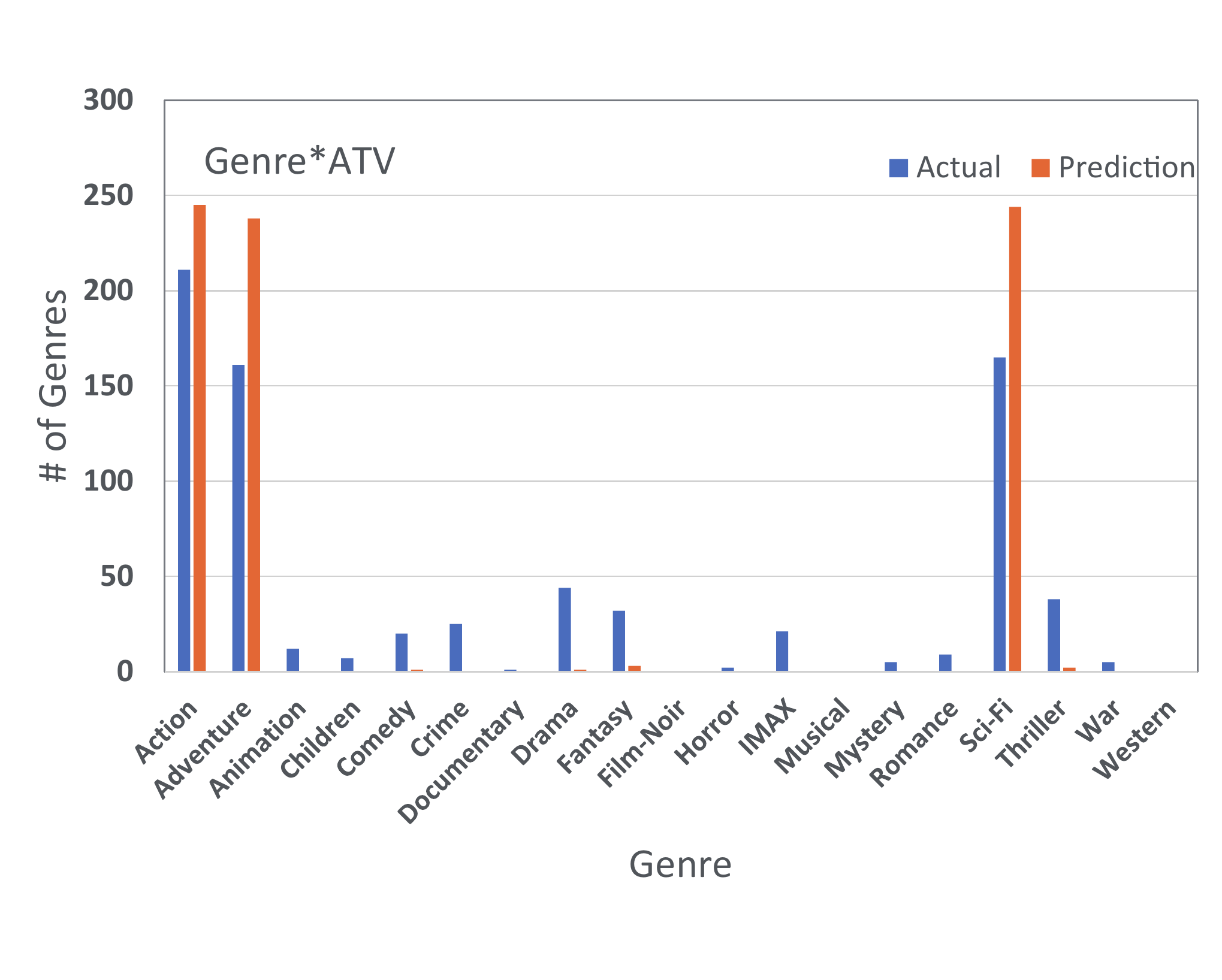}\label{fig:RNN2}}
\subfigure[~(RNN) Worst result among all clusters after clustering.]{\includegraphics[width=0.4\columnwidth]{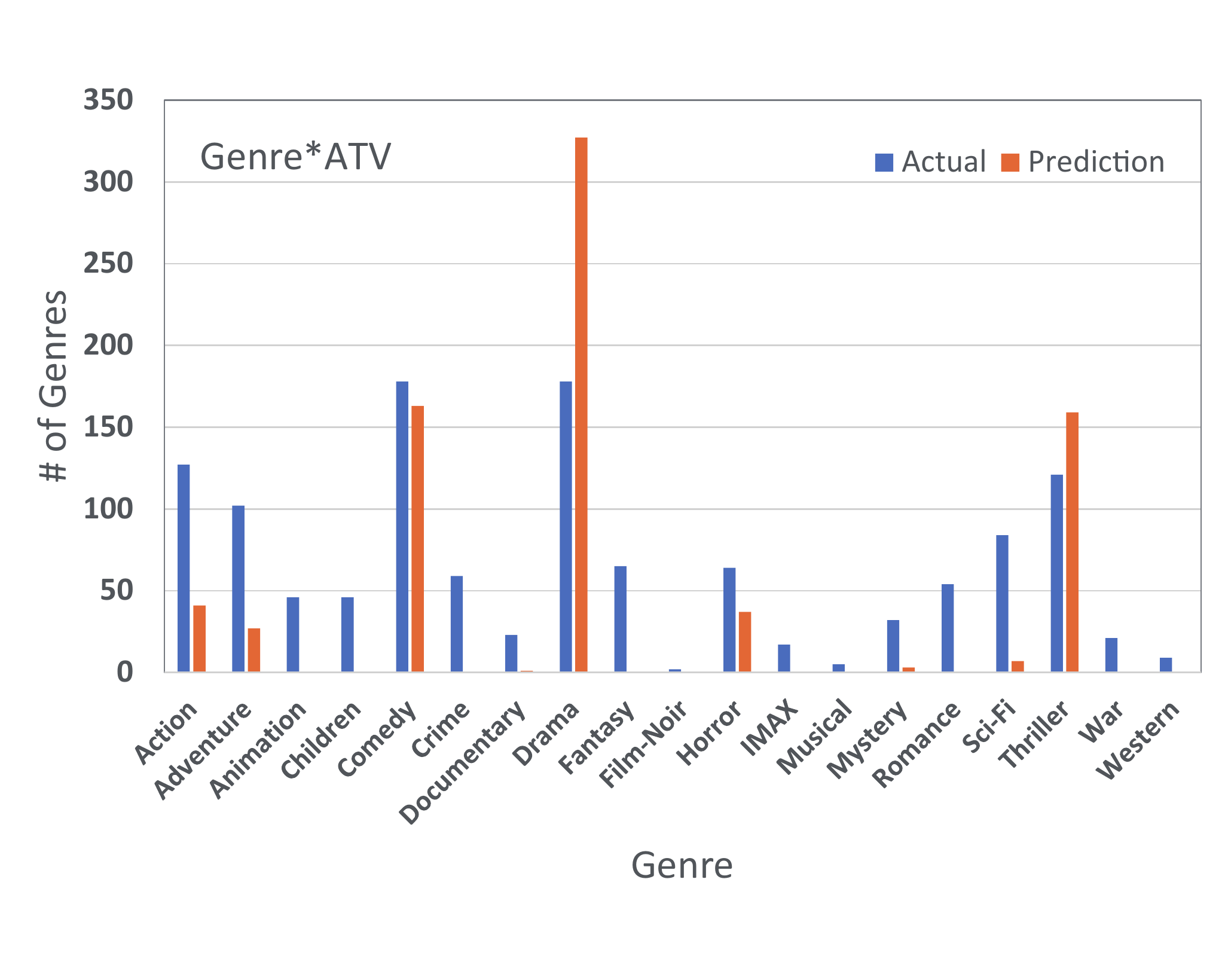}\label{fig:RNN3}}
\subfigure[~(LSTM) Result without clustering.]{\includegraphics[width=0.4\columnwidth]{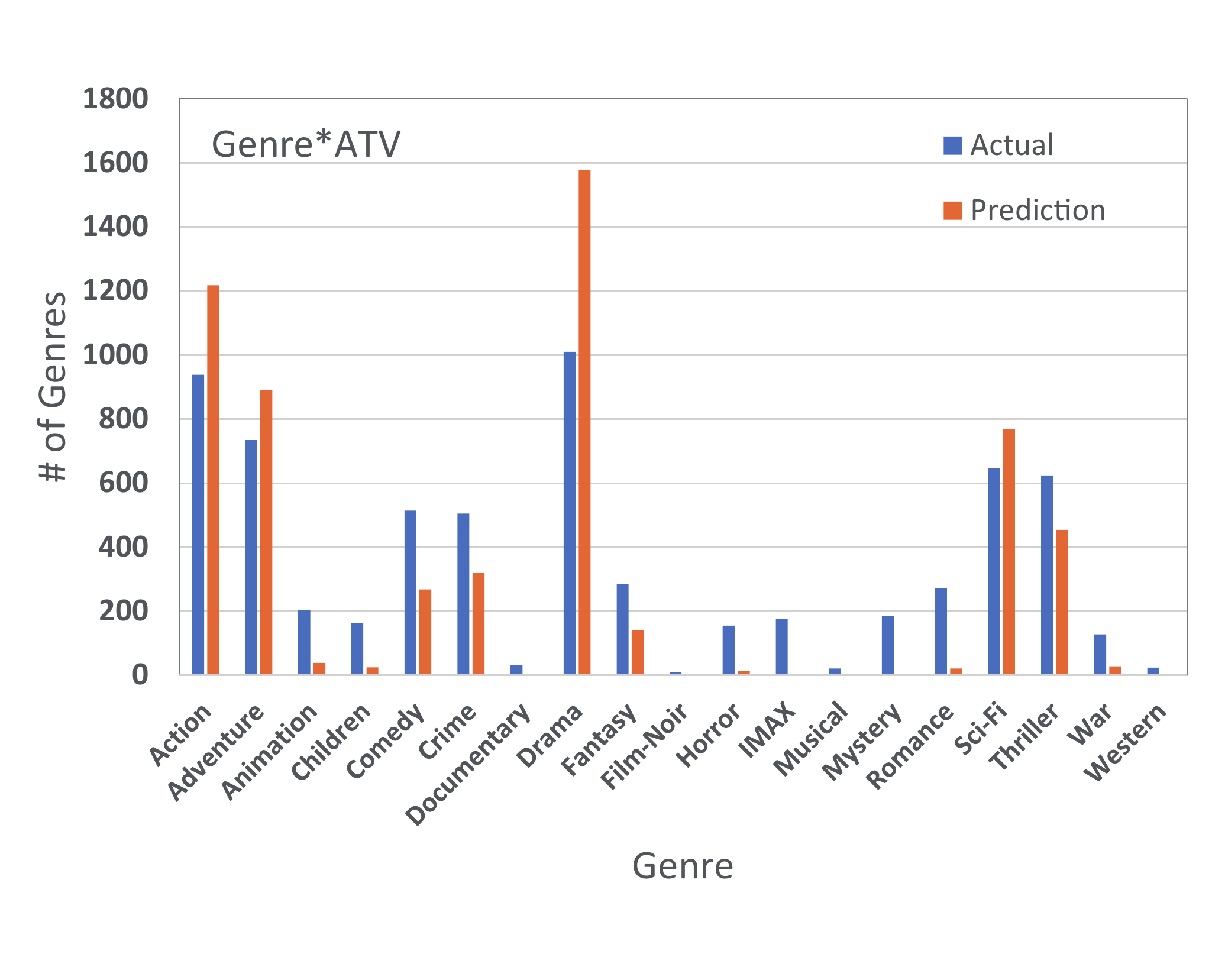}\label{fig:LSTM1}}
\subfigure[~(LSTM) Best result among all clusters after clustering.]{\includegraphics[width=0.4\columnwidth]{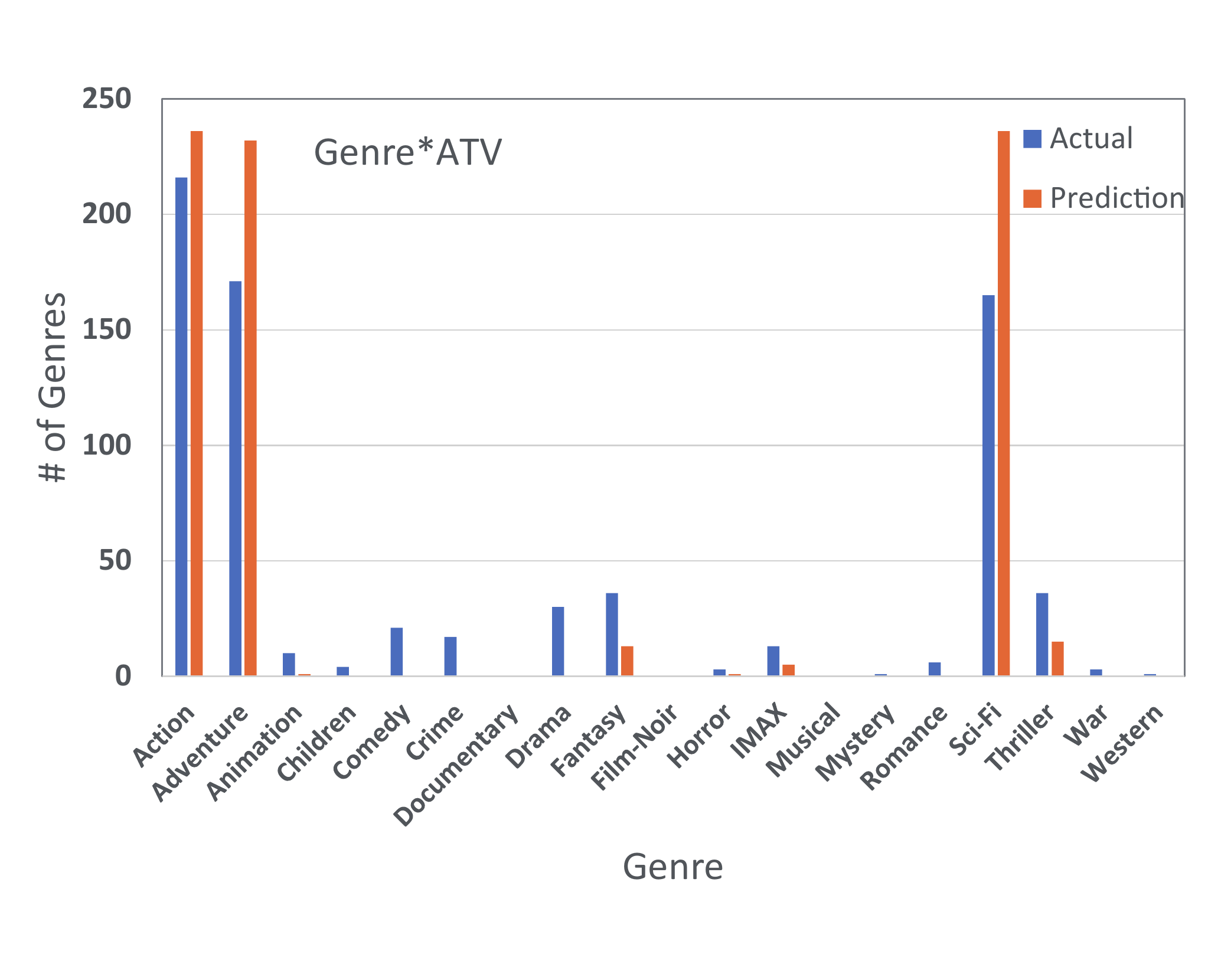}\label{fig:LSTM2}}
\subfigure[~(LSTM) Worst result among all clusters after clustering.]{\includegraphics[width=0.4\columnwidth]{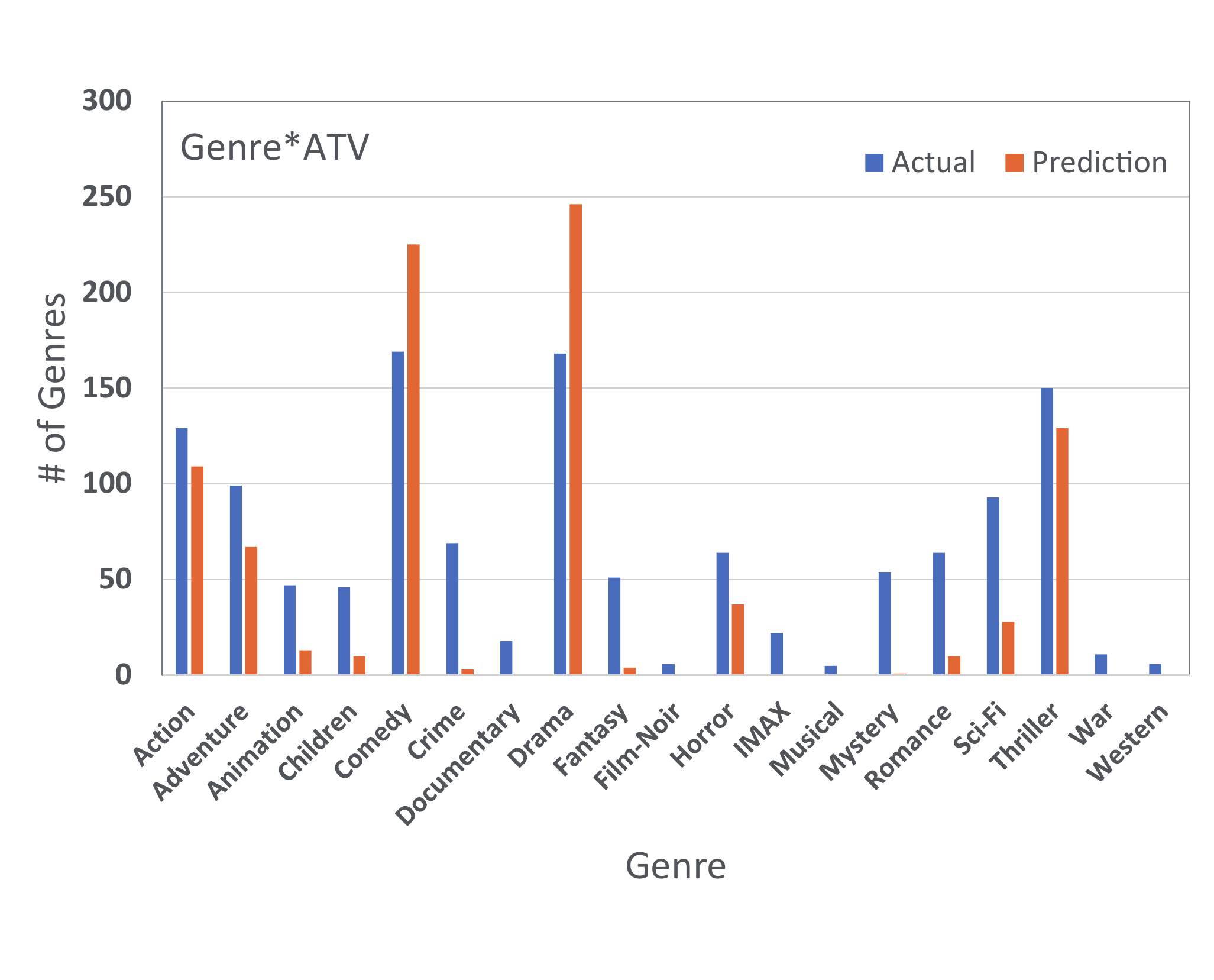}\label{fig:LSTM3}}
\subfigure[~(GRU) Result without clustering.]{\includegraphics[width=0.4\columnwidth]{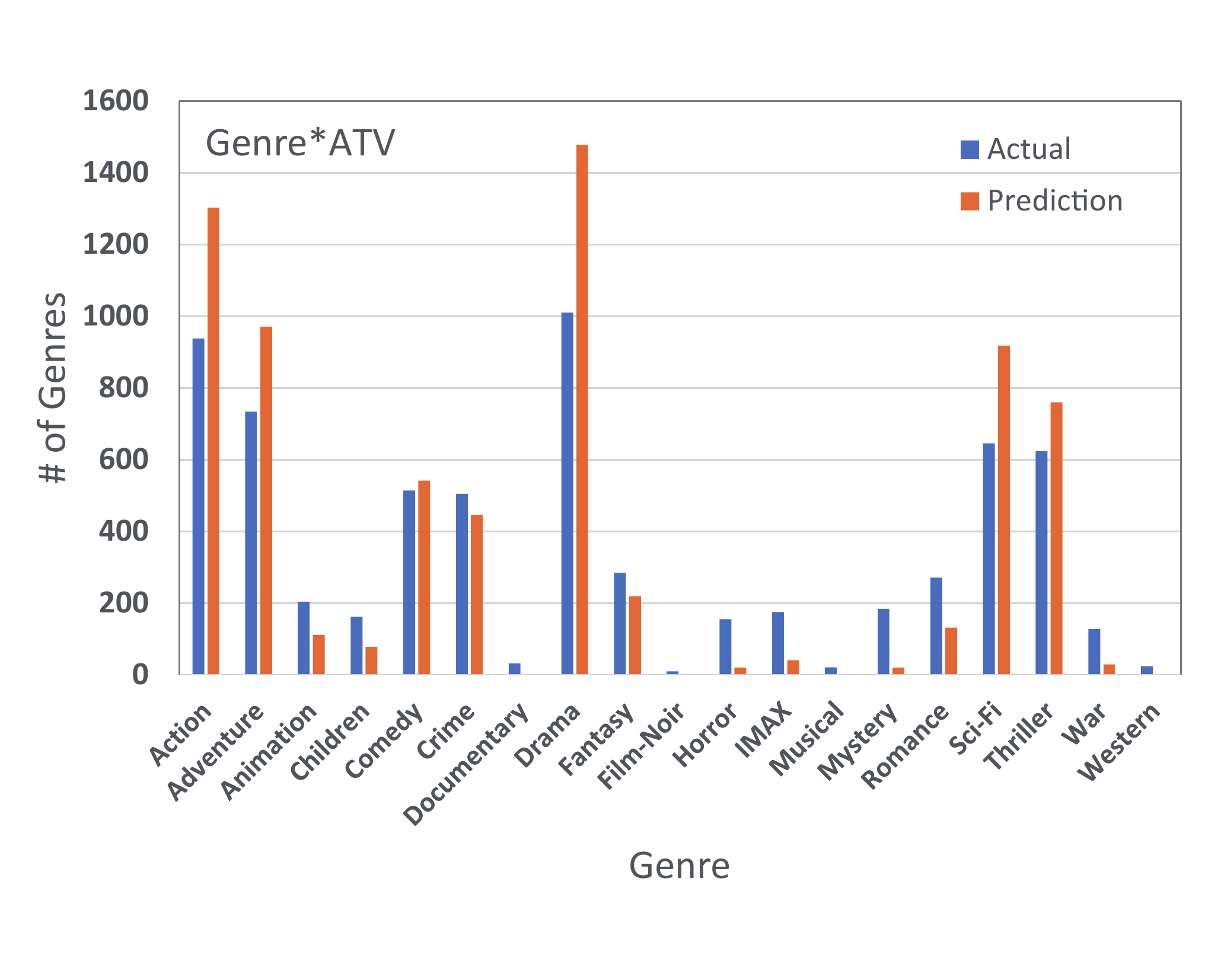}\label{fig:GRU1}}
\subfigure[~(GRU) Best result among all clusters after clustering.]{\includegraphics[width=0.4\columnwidth]{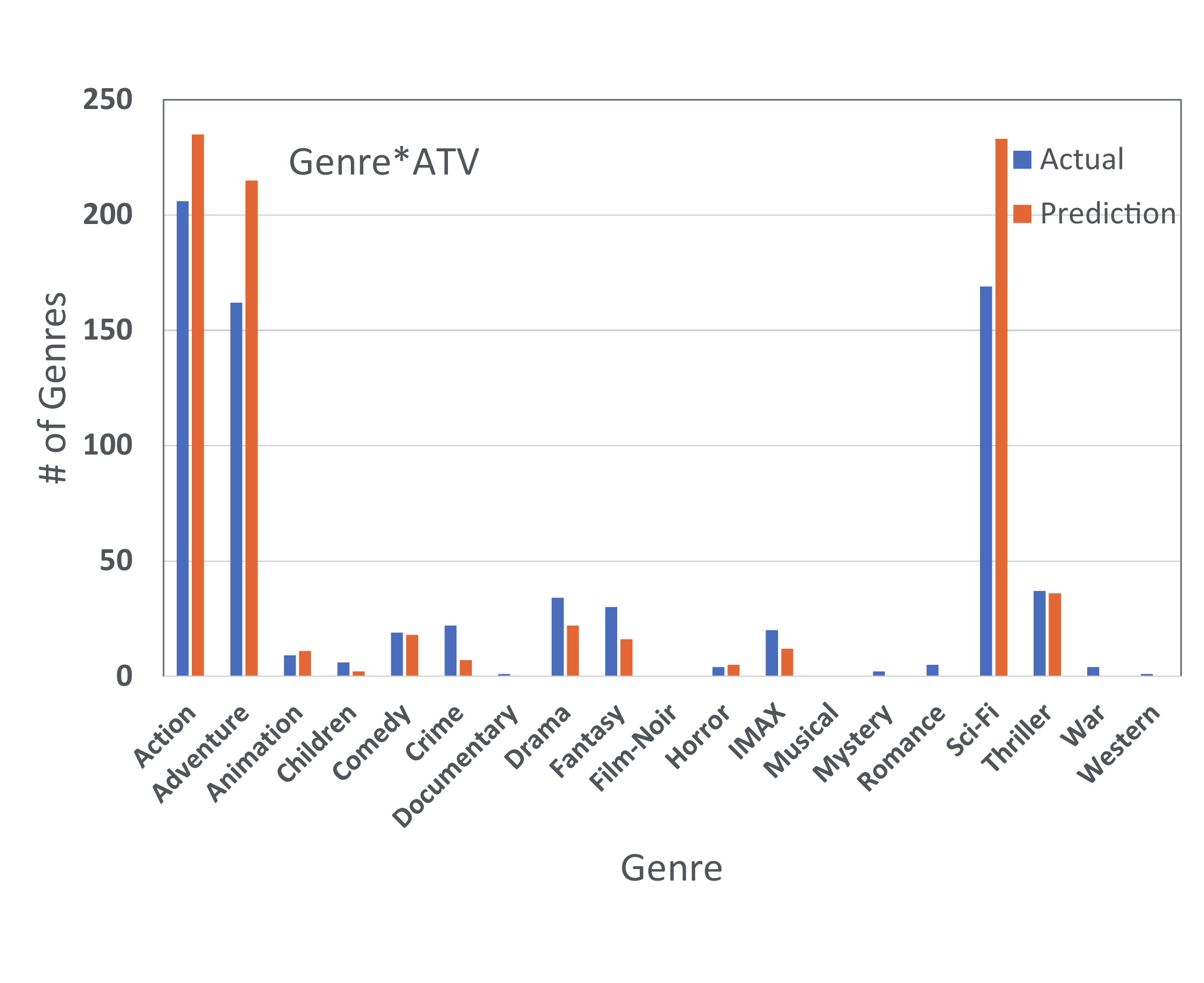}\label{fig:GRU2}}
%\subfigure[~(GRU) Worst result among all clusters after clustering.]{\includegraphics[width=0.3\columnwidth]{figs/GRU3}\label{fig:GRU3}}
\caption{Result of Clustering for RNN ((a), (b), (c)), LSTM ((d), (e), (f)), and GRU ((g), (h), (i)), respectively.}
\label{fig:clustering}
\vspace{-0.2cm}
\end{figure}

As a first result, we obtain the performances of each model before the clustering and after clustering in Figure~\ref{fig:clustering}. Without clustering (Figure~\ref{fig:RNN1}, Figure~\ref{fig:LSTM1} and Figure~\ref{fig:GRU1}), the performances are measured in consideration of all users without distinguishing users based on any criteria.
This is because the data is not classified, it is difficult for the models (RNN, LSTM and GRU) to grasp the data itself, and it is difficult to extract any information. Therefore, all three models show relatively poor performance. In order to improve performance, we apply clustering for users with similar preferences.
After the clustering (Figure~\ref{fig:RNN2}, Figure~\ref{fig:LSTM2} and Figure~\ref{fig:GRU2}), we see that the performance is quite improved, and among all stages of our experiment, the range of performance increase is large. In this case, users with similar preferences are grouped together, so that in the case of a group in which preferences are well expressed, the range of values between preferred and non-preferred genres is very large. In other words, the number of data from genres with clear preferences is overwhelmingly large. It can be said that this helped make the process of recommending movies that the group would like to be easier for the model. However, even after doing this, there were occasionally (1 or 2) clusters where the preference was not clearly evident (Figure~\ref{fig:RNN3}, Figure~\ref{fig:LSTM3} and Figure~\ref{fig:GRU3}). In Table~\ref{tab:RNN}-\ref{tab:GRU}, we obtain the results of four performance metrics (Recall, Precision, Accuracy and F1-score) with respect to three training models, respectively. In the experiment, we consider the training data as [Genre*ATV] as a representative one. Here, the abbreviations BC means before clustering and AC means after clustering. AC (best) and AC (worst) indicate the best result and worst result among clusters. Finally, AC (mean) present a mean of all result of clusters.
As a result, we see that the performances of AC for four metrics are improved compared to BC except for the worst case. Further, among them, we also check that the accuracy has the highest value since false positives and false negatives of the datasets are not quite different.
\begin{figure}[H]
\centering
\subfigure[~(RNN) Result after trimming.]{\includegraphics[width=0.45\columnwidth]{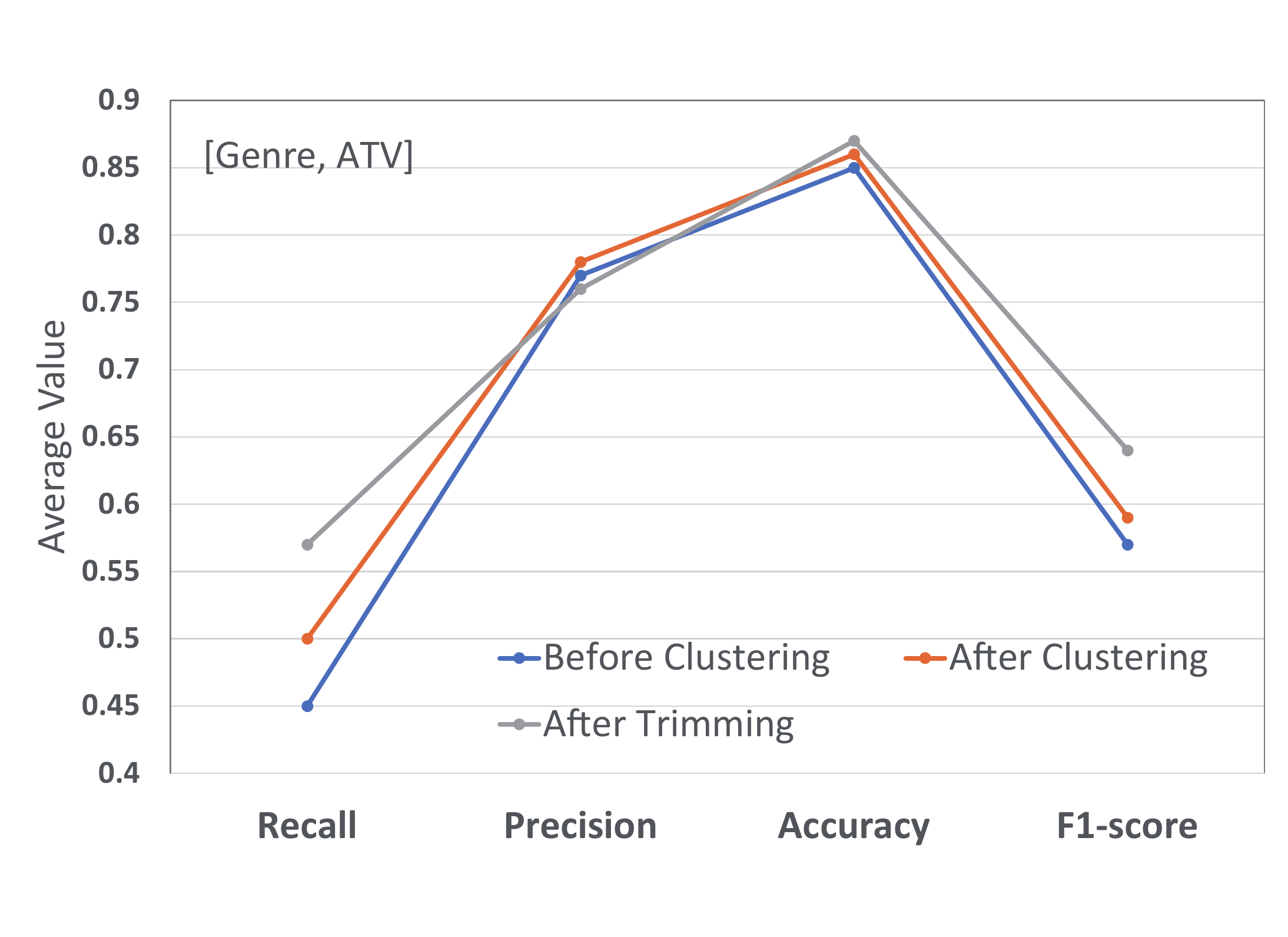}\label{fig:TrimmingRNN1}}
\subfigure[~(RNN) Result after trimming.]{\includegraphics[width=0.45\columnwidth]{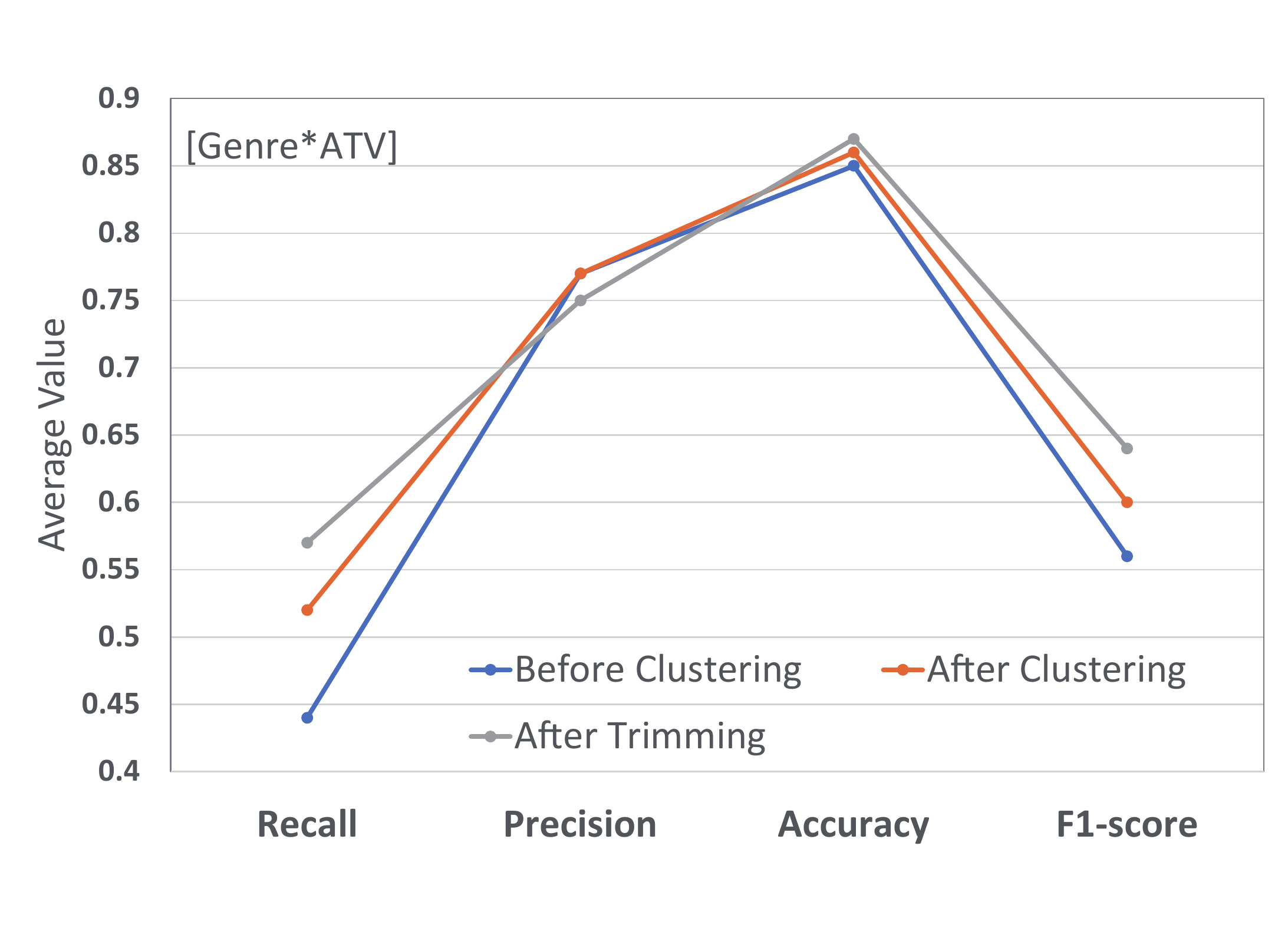}\label{fig:TrimmingRNN2}}
\subfigure[~(LSTM) Result after trimming.]{\includegraphics[width=0.45\columnwidth]{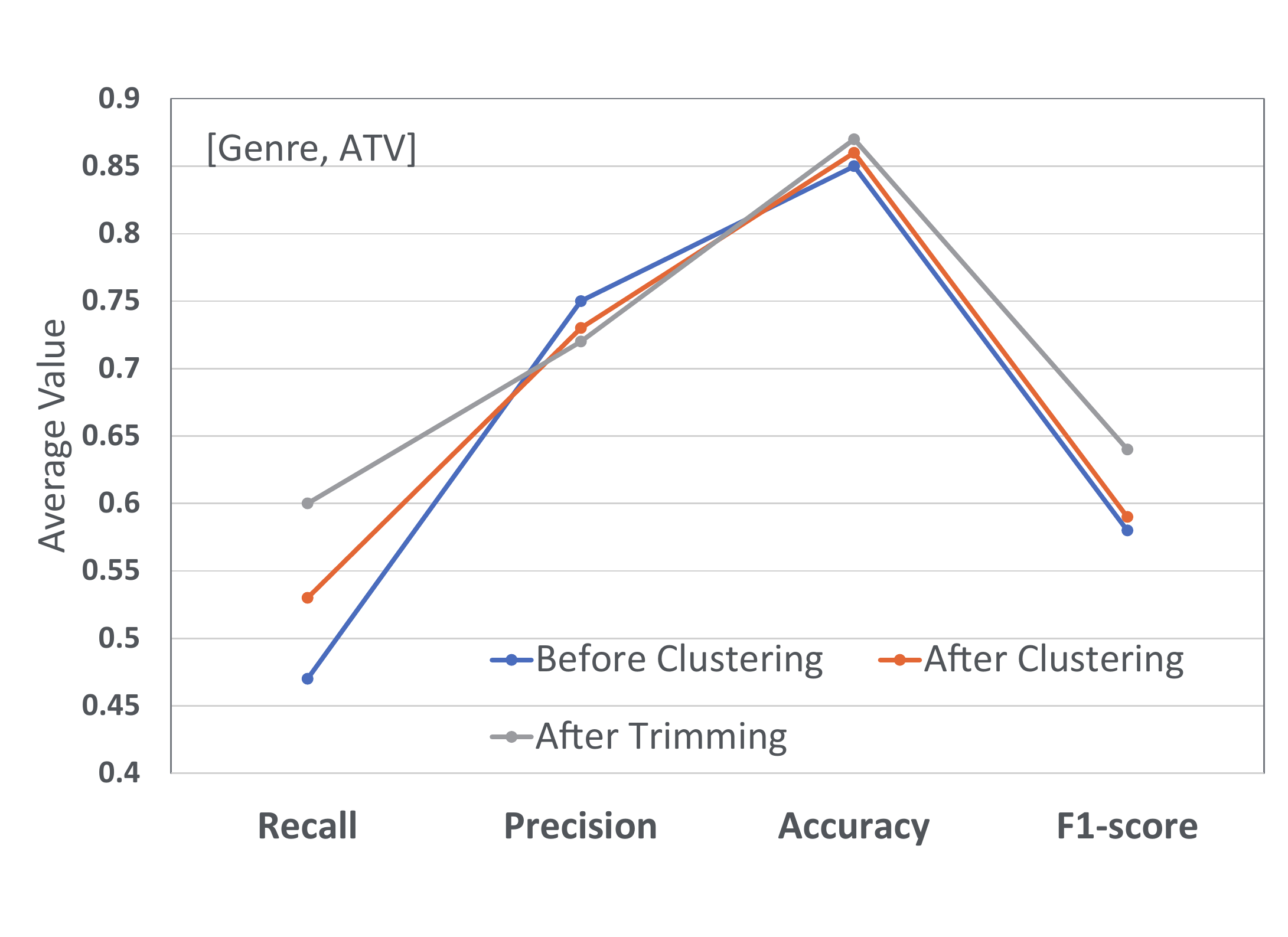}\label{fig:TrimmingLSTM1}}
\subfigure[~(LSTM) Result after trimming.]{\includegraphics[width=0.45\columnwidth]{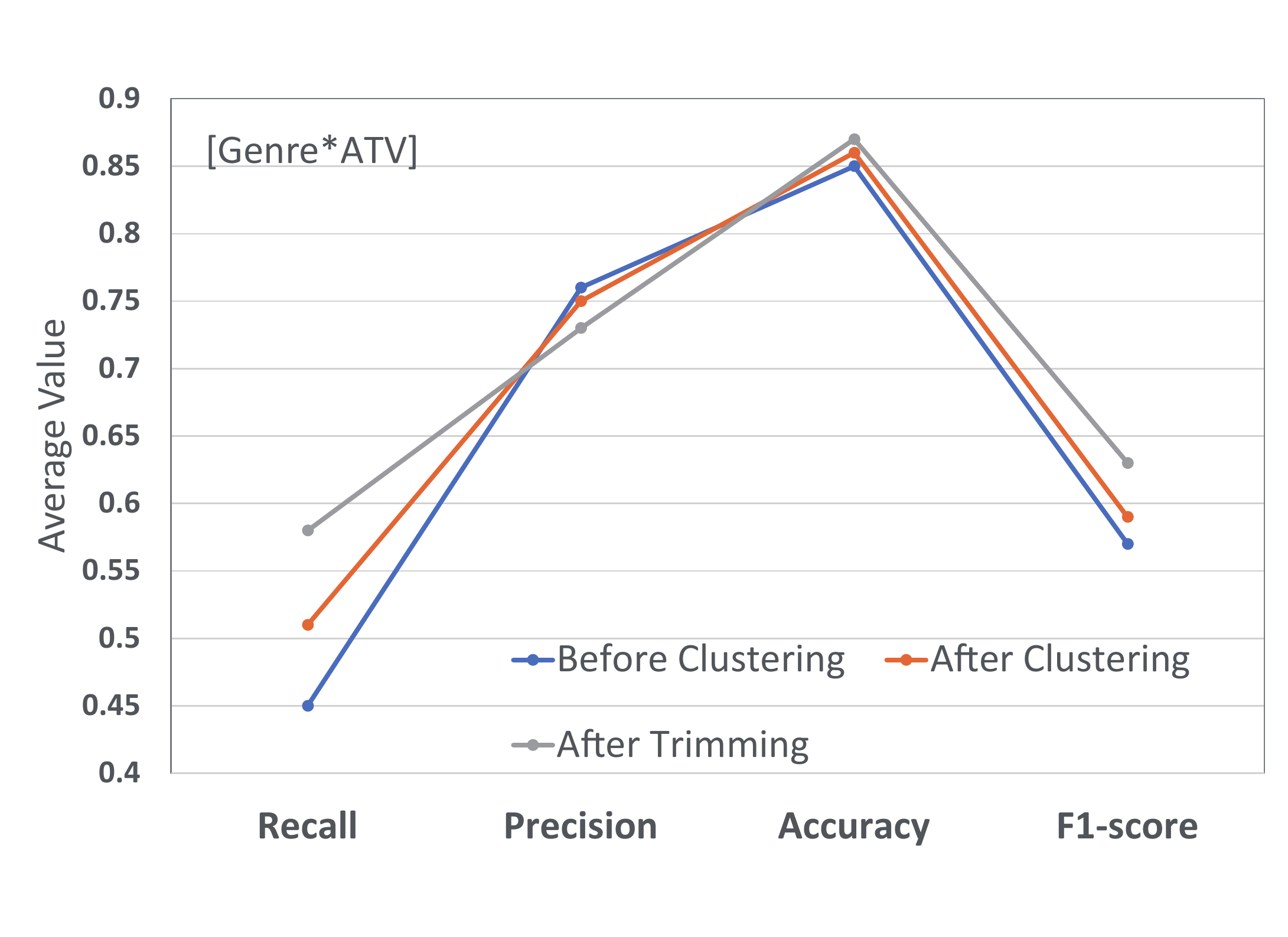}\label{fig:TrimmingLSTM2}}
\subfigure[~(GRU) Result after trimming.]{\includegraphics[width=0.45\columnwidth]{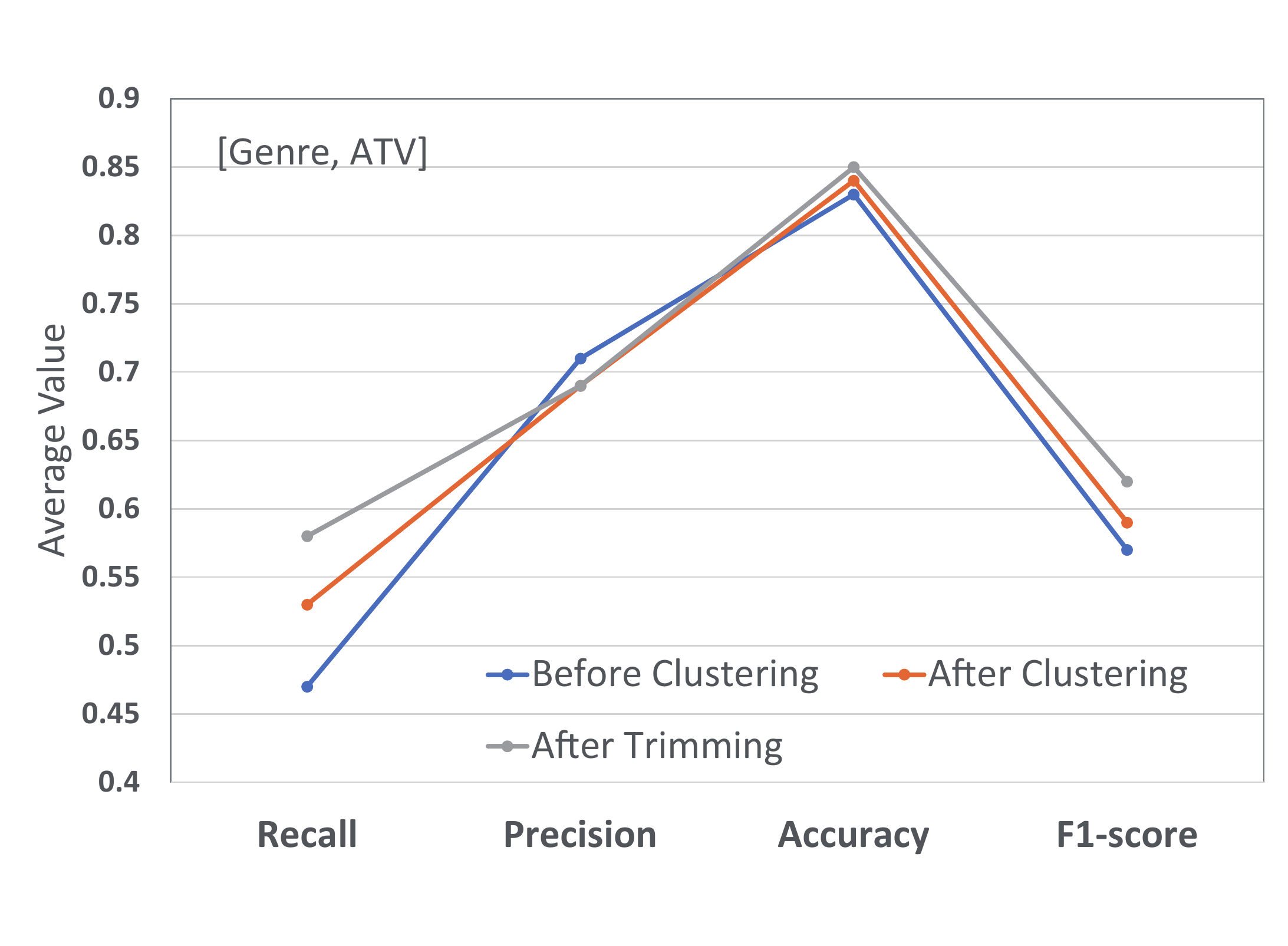}\label{fig:TrimmingGRU1}}
\subfigure[~(GRU) Result after trimming.]{\includegraphics[width=0.45\columnwidth]{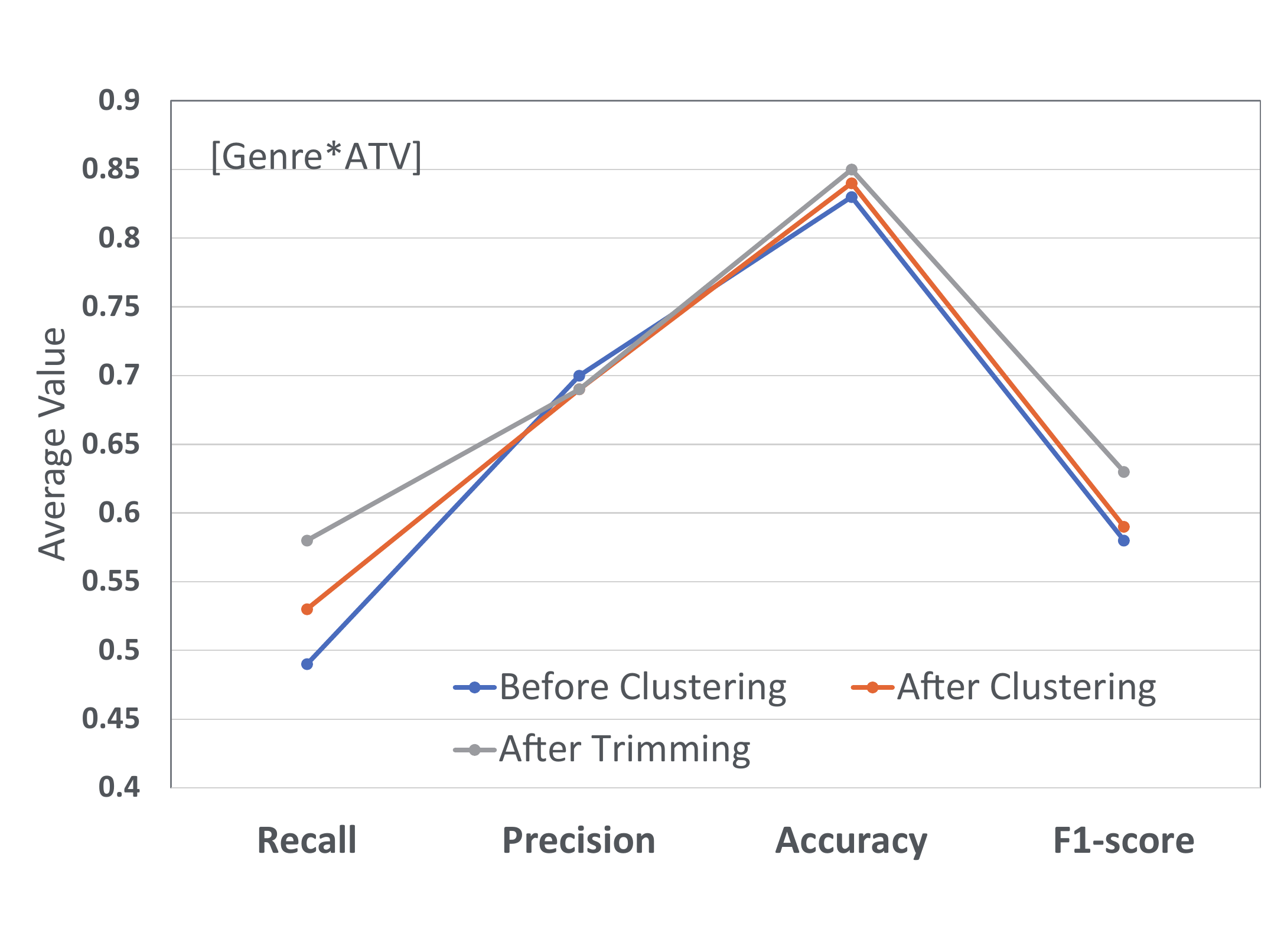}\label{fig:TrimmingGRU2}}
\caption{Result of Sub-Genre Trimming for RNN ((a), (b)), LSTM ((c), (d)), and GRU ((e), (f)), respectively.}
\label{fig:trimming}
\vspace{-0.2cm}
\end{figure}

% \begin{figure*}[t!]
% \centering
% \subfigure[~Recovering vs. Observation probability.]{\includegraphics[width=0.6\columnwidth]{figs/TrimmingLSTM1}\label{fig:TrimmingLSTM1}}
% \subfigure[~Recovering vs. Mean \# of infected nodes]{\includegraphics[width=0.6\columnwidth]{figs/TrimmingLSTM2}\label{fig:TrimmingLSTM2}}
% \caption{Result of Trimming LSTM}
% \label{fig:LSTM}
% \vspace{-0.2cm}
% \end{figure*}

% \begin{figure*}[t!]
% \centering
% \subfigure[~Recovering vs. Observation probability.]{\includegraphics[width=0.6\columnwidth]{figs/TrimmingGRU1}\label{fig:TrimmingGRU1}}
% \subfigure[~Recovering vs. Mean \# of infected nodes]{\includegraphics[width=0.6\columnwidth]{figs/TrimmingGRU2}\label{fig:TrimmingGRU2}}
% \caption{Result of Trimming GRU}
% \label{fig:LSTM}
% \vspace{-0.2cm}
% \end{figure*}

\begin{table}[H]
\caption{Four performance results of RNN with the training data type [Genre*ATV] }
\label{tab:RNN1}
\centering
 \begin{tabular}{c|cccc}
\hline  Trimming   & Recall & Precision & Accuracy & F1-score \\
\hline BT (mean) &  0.52 %Is the bold necessary?
 & 0.77&0.86&0.60\\
\hline BT (worst)& 0.29 & 0.87&0.85&0.43\\
\hline AT (worst)& 0.37 & 0.87&0.85&0.52\\
\hline AT (mean)& 0.57 & 0.75 & 0.87&0.64\\
\hline
\end{tabular} \end{table}

\begin{table}[H]
\caption{Four performance results of LSTM with the training data type [Genre*ATV] }
\label{tab:LSTM1}
\centering
 \begin{tabular}{c|cccc}
\hline  Clustering    & Recall & Precision & Accuracy & F1-score \\
\hline BT (mean) &  0.51 %Is the bold necessary?
 & 0.75&0.86&0.59\\
\hline BT (worst)& 0.27 & 0.84&0.84&0.41\\
\hline AT (worst)& 0.35 & 0.84&0.86&0.49\\
\hline AT (mean)& 0.58 & 0.73 & 0.87&0.63\\
\hline
\end{tabular} \end{table}

\begin{table}[H]
\caption{Four performance results of GRU with the training data type [Genre*ATV] }
\label{tab:GRU1}
\centering
 \begin{tabular}{c|cccc}
\hline  Clustering    & Recall & Precision & Accuracy & F1-score \\
\hline BT (mean) &  0.53 %Is the bold necessary?
 & 0.69&0.84&0.59\\
\hline BT (worst)& 0.35 & 0.69&0.79&0.46\\
\hline AT (worst)& 0.43 & 0.70&0.82&0.53\\
\hline AT (mean)& 0.58 & 0.69 & 0.85&0.63\\
\hline
\end{tabular} \end{table}

\subsubsection{Effects of Sub-genre Trimming}
In order to maximize the advantages of clustering, we come up with a method of trimming the sub-genres that are not preferred in the cluster. To see this, we set the threshold $\eta$ by 0.5 and if there is a cluster that does not exceed 0.5 at any one of Recall, Precision, Accuracy and F1-score, it is subject to trimming. For the result, we consider the following three cases: before clustering, after clustering, and after trimming. We use two kinds of training data types such as [Genre, ATV] and [Genre*ATV] for each model. As shown in Figure~~\ref{fig:trimming}, we see that most of the metrics for the after trimming case has larger value than others except precision. This is because the precision considers the ratio of correctly predicted positive observations to the total predicted positive observations. Further, we also check that the accuracy is the highest values for all three models.
In Table~\ref{tab:RNN1}-\ref{tab:GRU1}, we obtain the results of four performance metrics (Recall, Precision, Accuracy and F1-score) with respect to three training models, respectively. In the experiment, we consider the training data as [Genre*ATV] as a representative one. Here, the abbreviations BT means before trimming, and AT means after trimming. AT (best) and AT (worst) indicate the best result and worst result among clusters. Finally AT (mean) present a mean of all result of clusters.
As a result, we also check that the performances of AT for four metrics are improved.

\begin{figure}[H]
\centering
\subfigure[~(RNN) Result of ATV.]{\includegraphics[width=0.45\columnwidth]{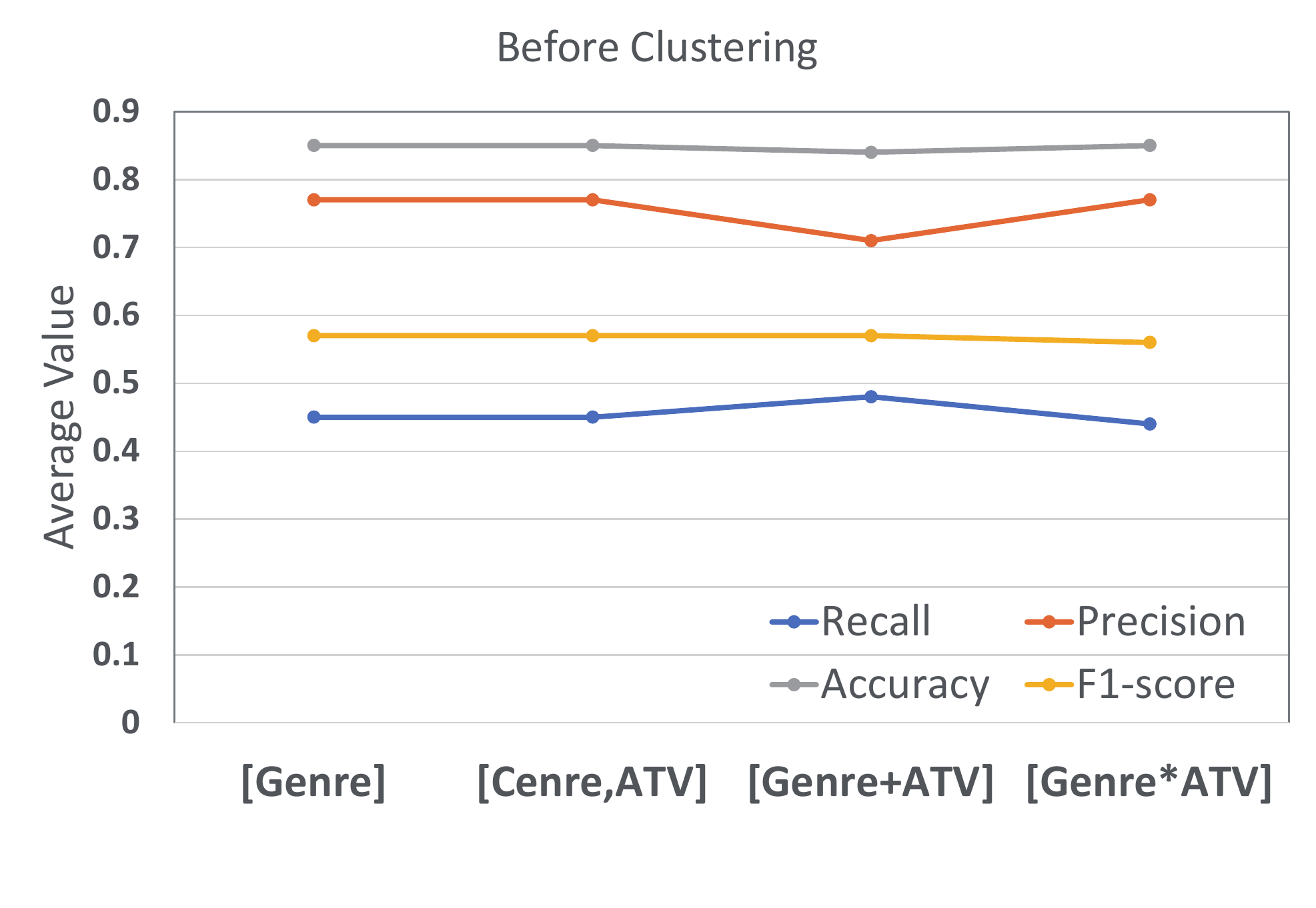}\label{fig:TVRNN1}}
\subfigure[~(RNN) Result of ATV.]{\includegraphics[width=0.45\columnwidth]{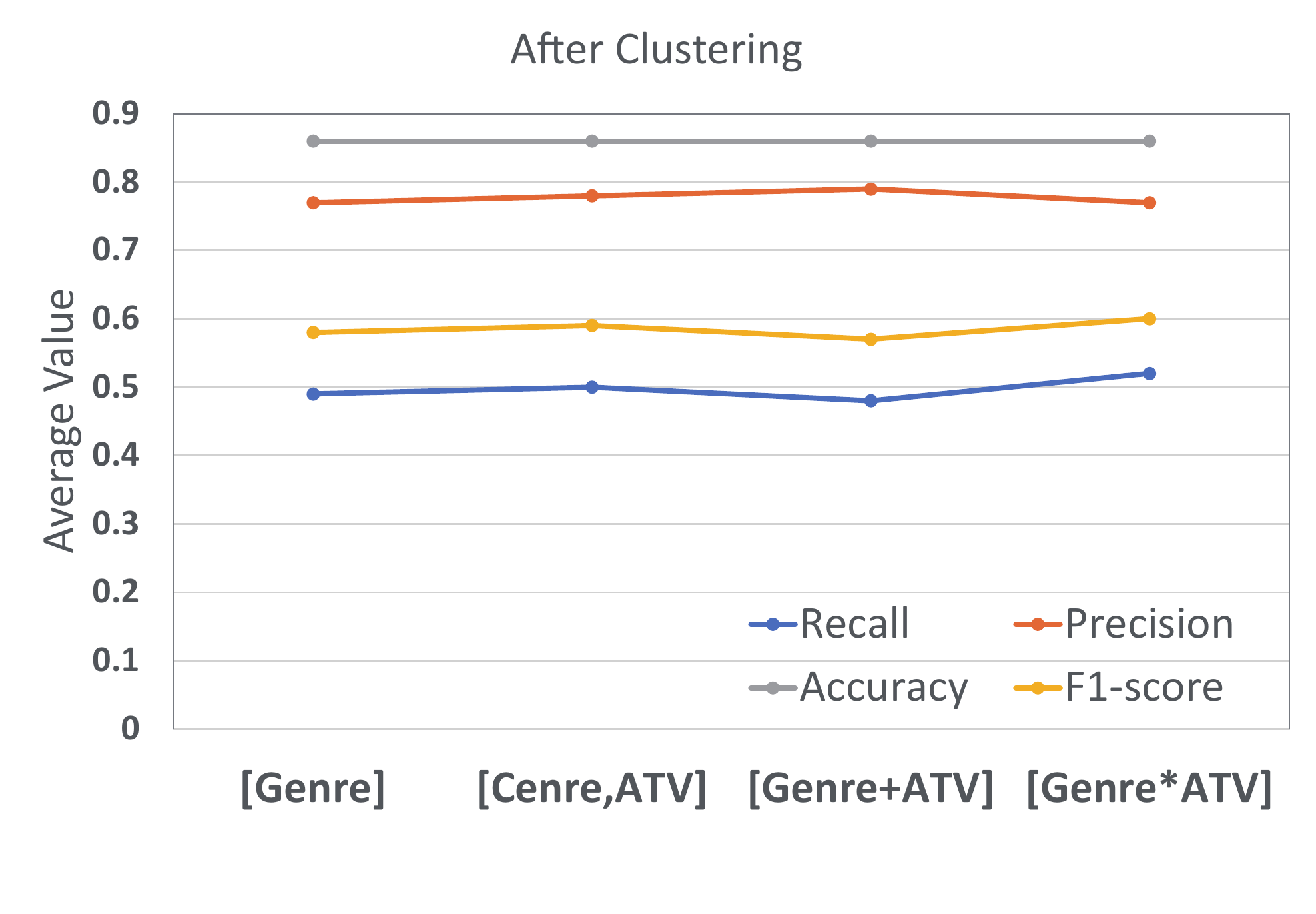}\label{fig:TVRNN2}}
\subfigure[~(RNN) Result of ATV.]{\includegraphics[width=0.45\columnwidth]{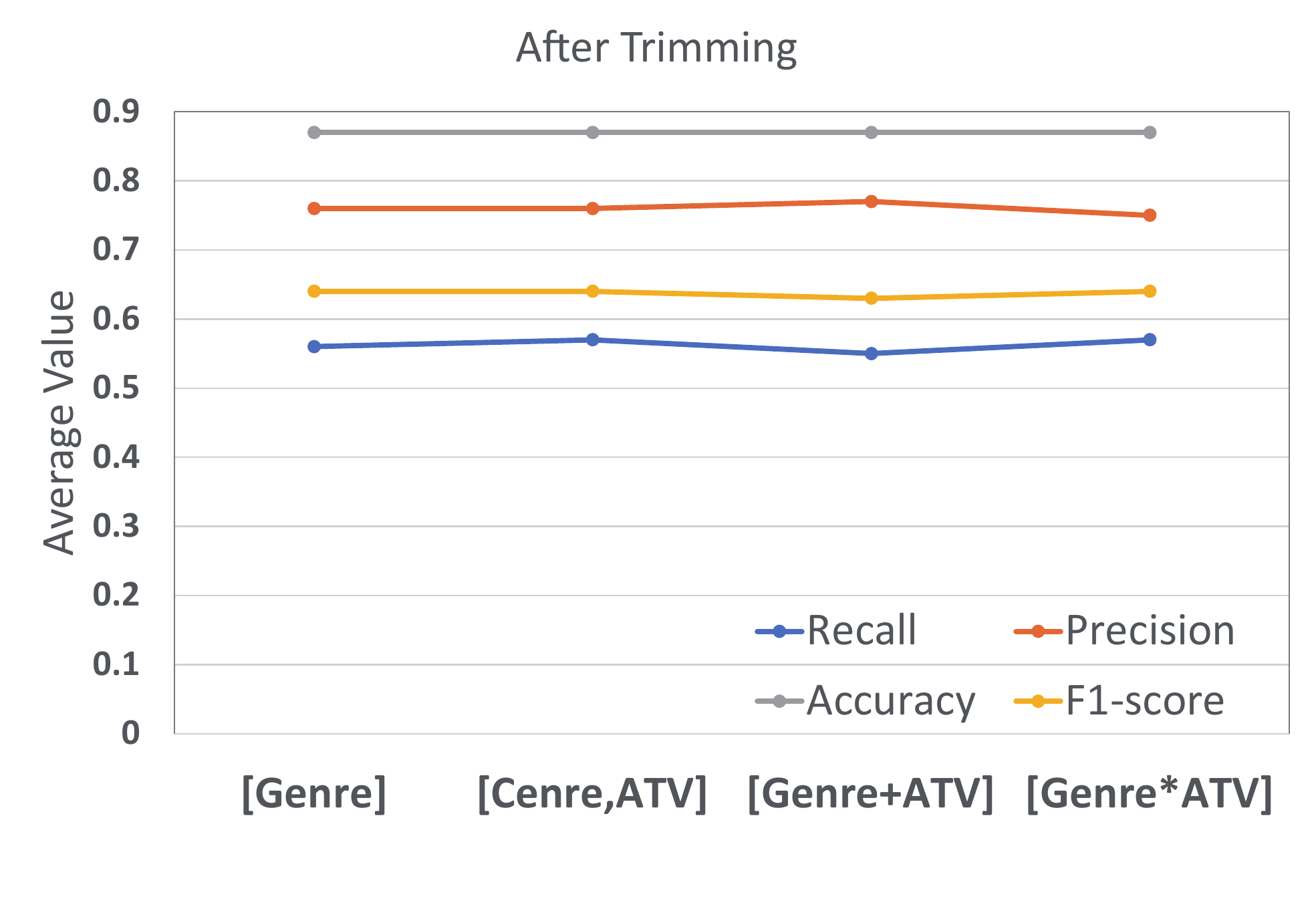}\label{fig:TVRNN3}}
\subfigure[~(LSTM) Result of ATV.]{\includegraphics[width=0.45\columnwidth]{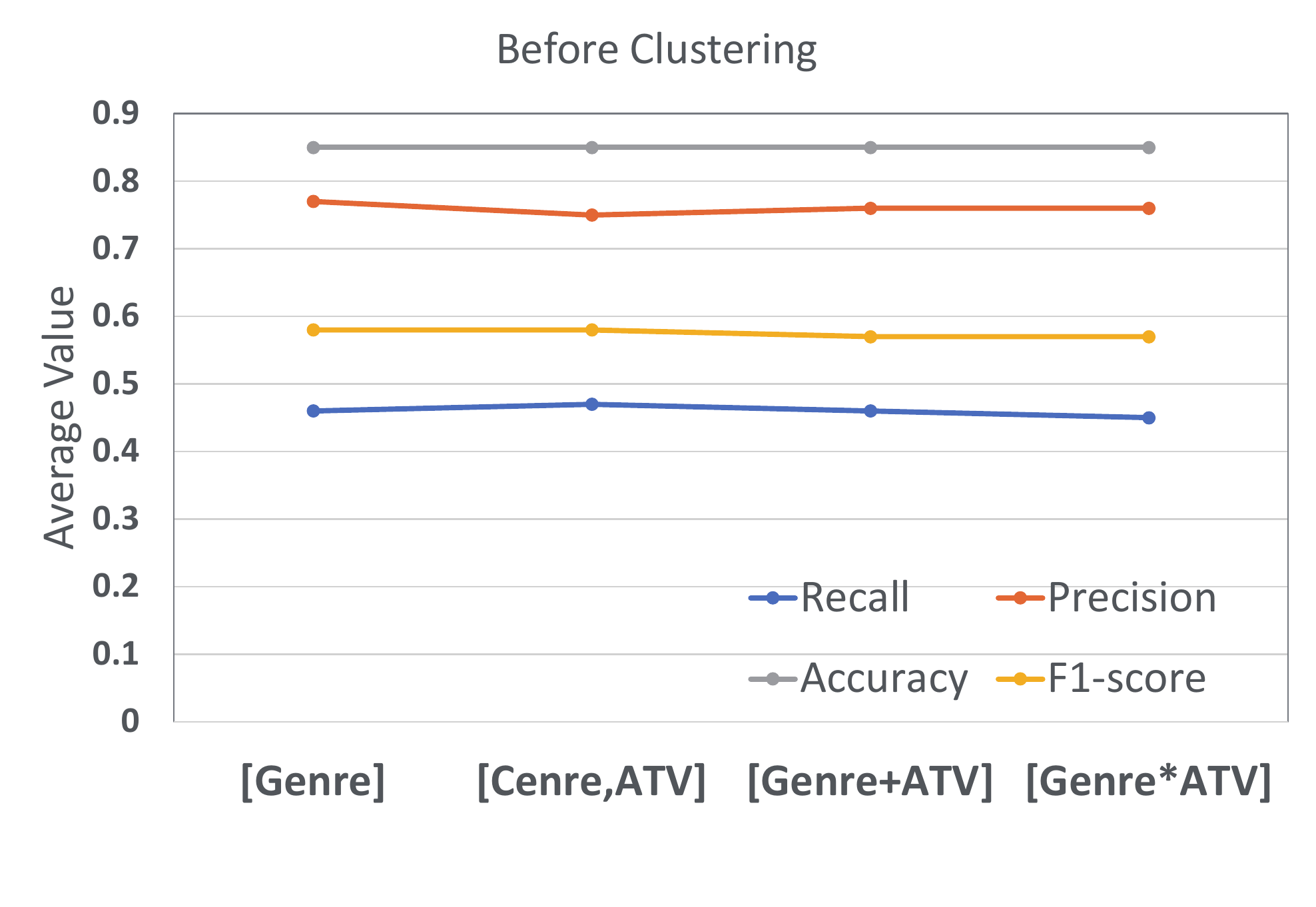}\label{fig:TVLSTM1}}
\subfigure[~(LSTM) Result of ATV.]{\includegraphics[width=0.45\columnwidth]{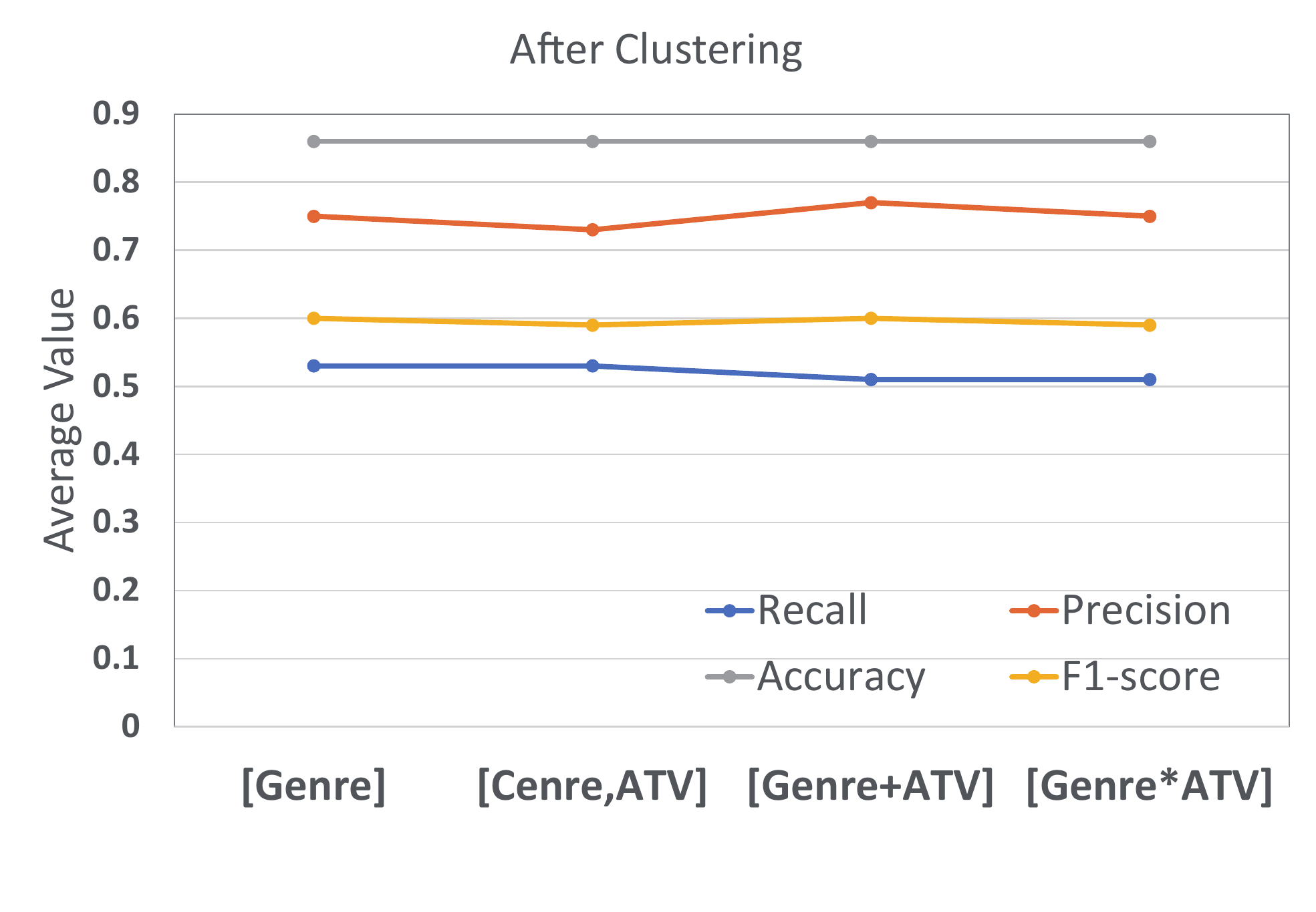}\label{fig:TVLSTM2}}
\subfigure[~(LSTM) Result of ATV.]{\includegraphics[width=0.45\columnwidth]{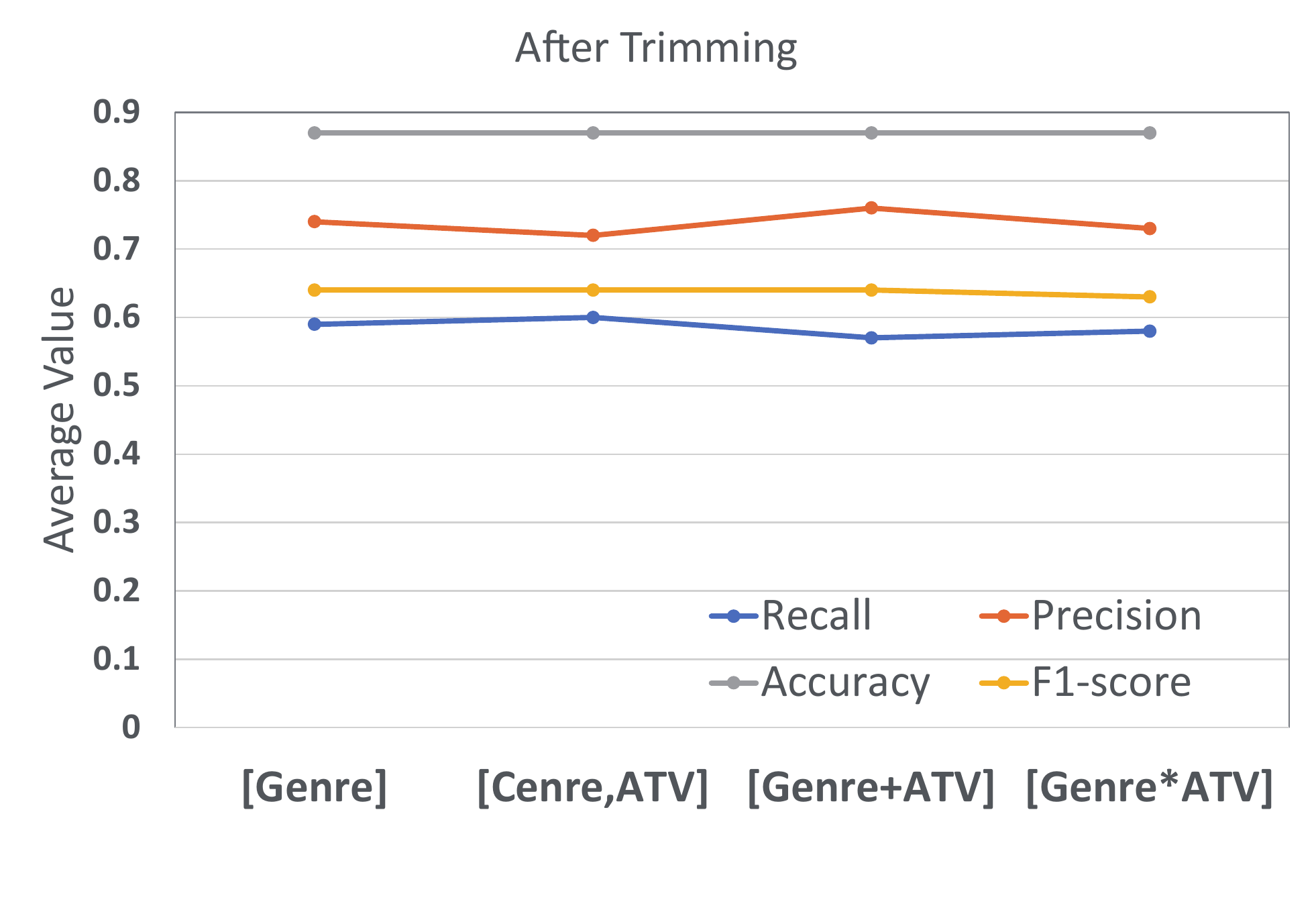}\label{fig:TVLSTM3}}
\subfigure[~(GRU) Result of ATV.]{\includegraphics[width=0.45\columnwidth]{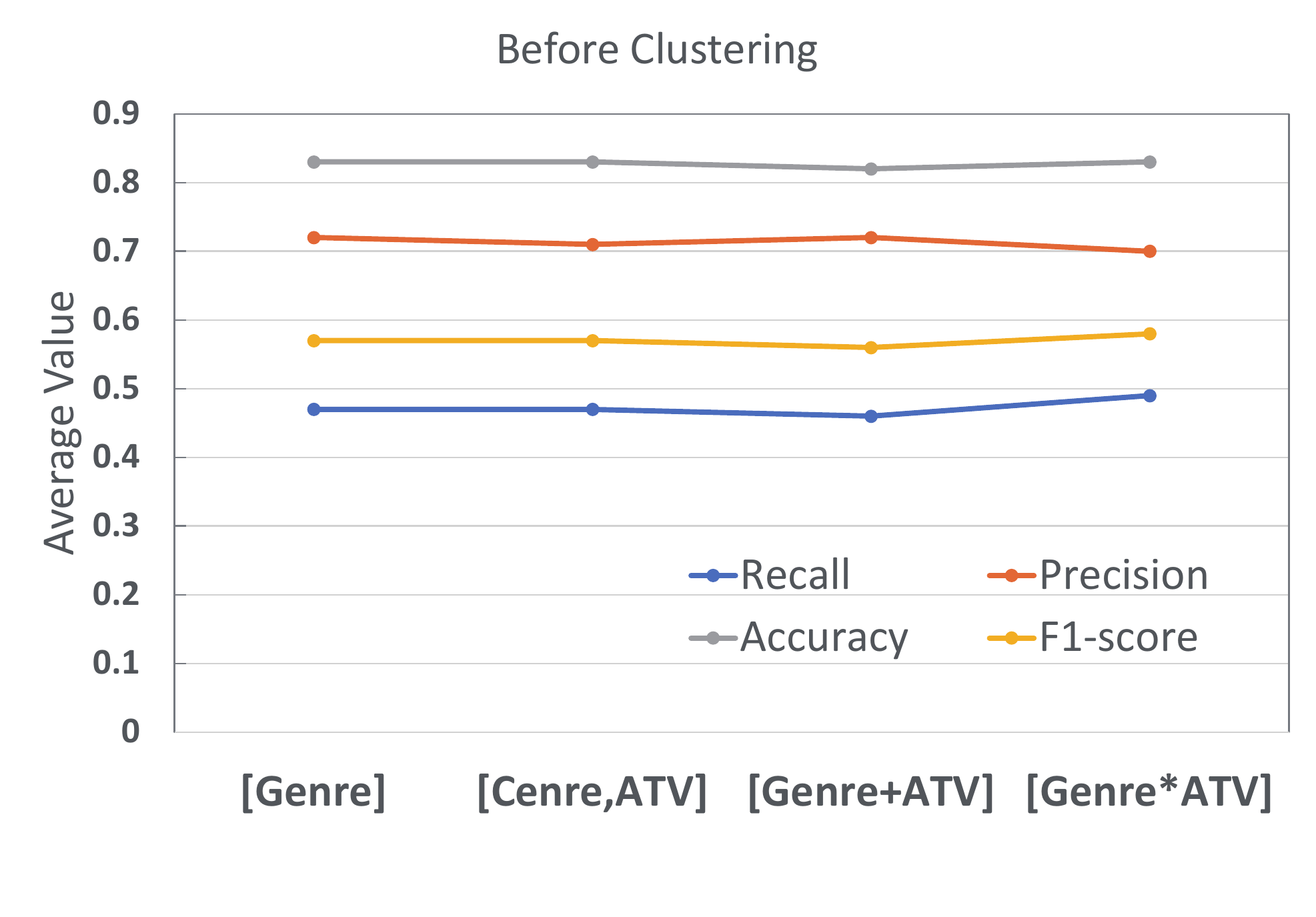}\label{fig:TVGRU1}}
\subfigure[~(GRU) Result of ATV.]{\includegraphics[width=0.45\columnwidth]{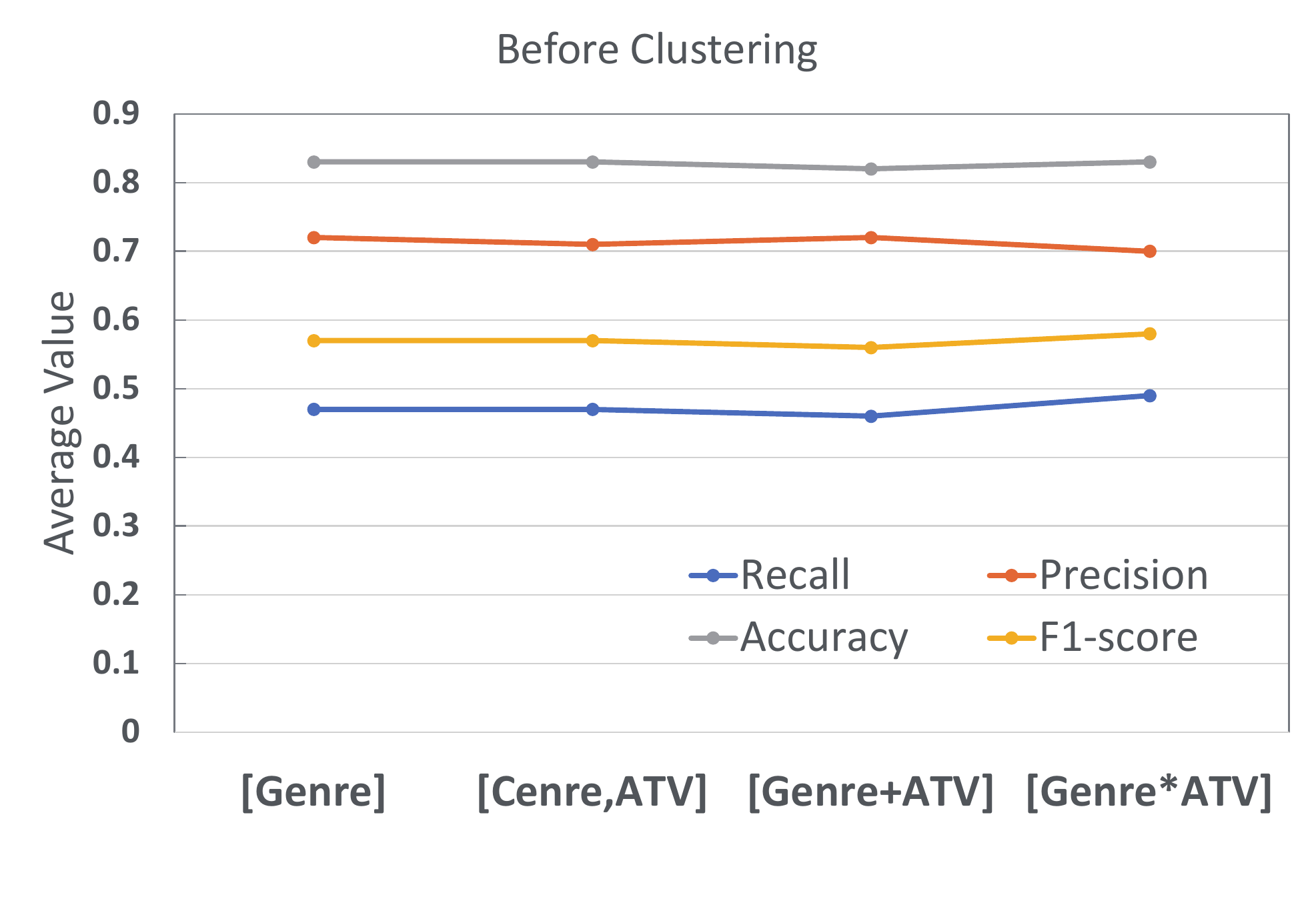}\label{fig:TVGRU1}}
\subfigure[~(GRU) Result of ATV.]{\includegraphics[width=0.45\columnwidth]{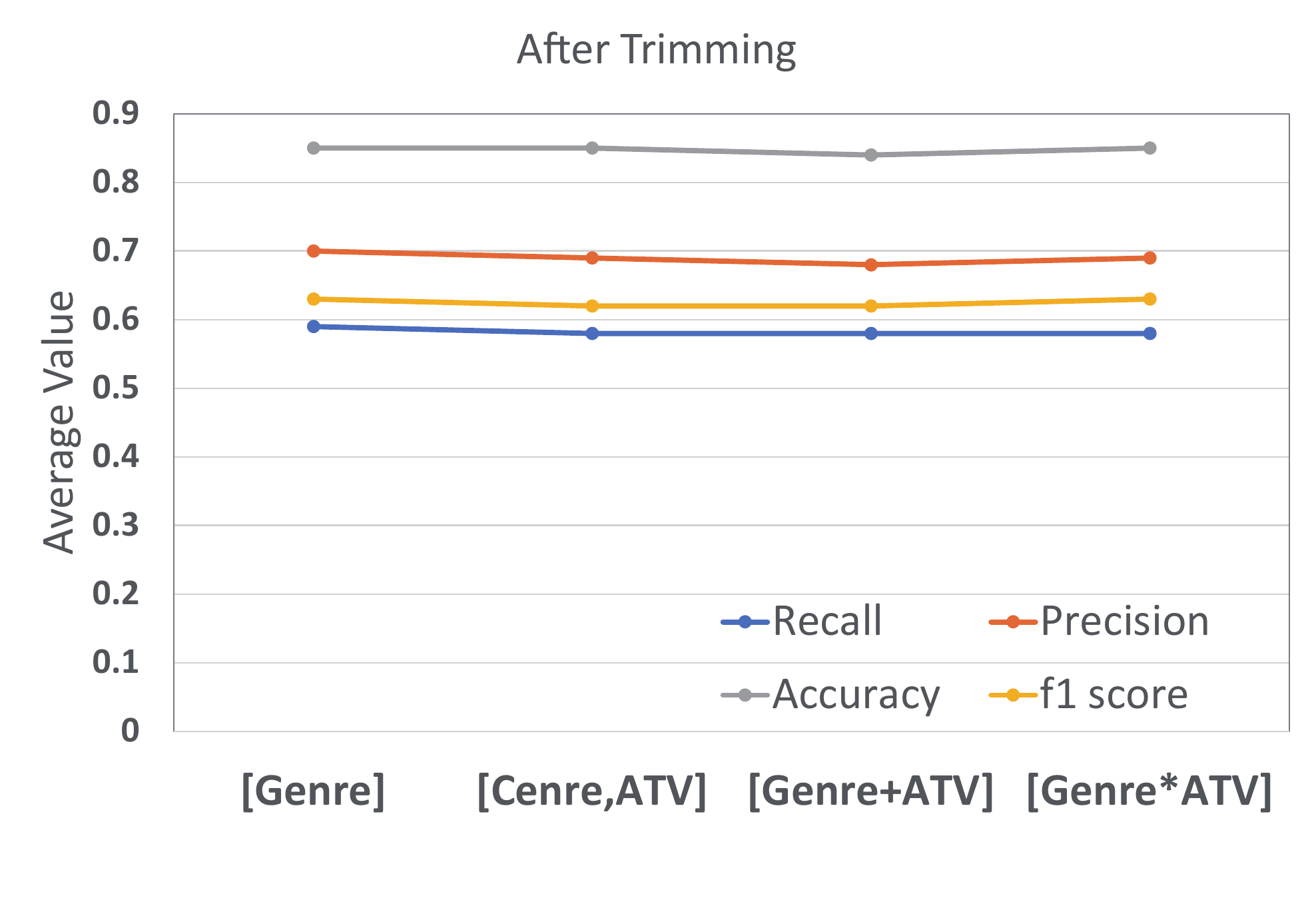}\label{fig:TVGRU1}}
\caption{Result of ATV for RNN((a), (b), (c)), LSTM ((d), (e), (f)), and GRU ((g), (h), (i)), respectively.}
\label{fig:ATV}
\vspace{-0.2cm}
\end{figure}

\subsubsection{Effects of Average Transition Probability}
Finally, we obtain recommended performance for the four data types described in the model to examine the effect of the ATV application. In the result, we consider the average evaluation values of all clusters.
As result, in Figure~\ref{fig:ATV}, we see that there was no significant difference in performance before and after clustering when the transition probability was applied and when the transition probability was not applied. Before clustering, there are no distinct characteristics to consider for all users, so it is seen as an environment that is not good for generating a transition probability that can contain the user's preference. After clustering, users with similar preferences are grouped together, so the ATV would have a good effect, but contrary to expectations, the effect is insignificant. Rather, except for RNN, the performance is slightly higher when the transition probability is excluded. Since RNN learns with more weight on recent information due to the characteristics of the model, it is expected that it will effectively represent the transition probability compared to other models. However,
the results show that the effects of clustering and trimming are still quite large.
% \begin{figure*}[t!]
% \centering
% \subfigure[~Recovering vs. Observation probability.]{\includegraphics[width=0.45\columnwidth]{figs/TVLSTM1}\label{fig:TVLSTM1}}
% \subfigure[~Recovering vs. Mean \# of infected nodes]{\includegraphics[width=0.45\columnwidth]{figs/TVLSTM2}\label{fig:TVLSTM2}}
% \subfigure[~Recovering vs. Parameter $\alpha$.]{\includegraphics[width=0.45\columnwidth]{figs/TVLSTM3}\label{fig:TVLSTM3}}
% \caption{Result of LSTM TV}
% \label{fig:LSTM}
% \vspace{-0.2cm}
% \end{figure*}

% \begin{figure*}[t!]
% \centering
% \subfigure[~Recovering vs. Observation probability.]{\includegraphics[width=0.45\columnwidth]{figs/TVGRU1}\label{fig:TVGRU1}}
% \subfigure[~Recovering vs. Mean \# of infected nodes]{\includegraphics[width=0.45\columnwidth]{figs/TVGRU2}\label{fig:TVGRU1}}
% \subfigure[~Recovering vs. Parameter $\alpha$.]{\includegraphics[width=0.45\columnwidth]{figs/TVGRU3}\label{fig:TVGRU1}}
% \caption{Result of GRU TV}
% \label{fig:LSTM}
% \vspace{-0.2cm}
% \end{figure*}

% detection probabilities for the simple batch querying scheme as
% varying $K$. As we expected, if $p$ is large, then the detection is
% better that of small $p$ for both graphs.
% In Fig.~\ref{fig:interactive_ersf}, we obtain the
% detection probabilities for the interactive querying with direction scheme as
% varying $K$. We see that the direction information is helpful to detect the source
% when $q>1/d$.

% \end{compactenum}
\section{Discussion}
\label{sec:discussion}
A method for genre prediction has been examined as a pre-step for sequential movie recommendation in this work. However, we did not specifically deal with how to make sequential movie recommendations based on this. In fact, in order to solve problems such as cold start due to the limitation of movie data in movie recommendation, \cite{Choi2012} also proposed a recommendation system based on correlation information on movie genres. However, this study does not suggest a recommendation method for movies with sequential dynamics. In the sequential movie recommendation system, it is necessary to study how to use the information on the genre of a movie to show the prediction performance well. In addition, it is necessary to design how to select and recommend movies including recommended genres based on the results of sequential movie genre prediction shown in our study.
As a learning method for short-term dynamics, we used ATV in Markov Chain. However, the reason why it did not have much effect on the performance is because the RNN-based deep learning we used actually learns some short-term dynamics. Therefore, we will examine whether this short-term dynamic is better estimated by additionally using a higher-order MC that uses information from the past better than the first-order MC that uses the estimation of the next step with the result of the previous step also need. We remain these as our future work.

\section{Conclusion}
\label{sec:conclusion}
In this paper, we proposed a sequential movie genre prediction algorithm based on the MC for the short-term behavior and RNN for the long-term behavior of user preference. The movie genre prediction does not recommend a specific movie, but it recommends the genre for the next movie to watch in consideration of each user's preference for the movie genre based on the genre included in the movie. For this, we considered that users with similar genre preferences are organized into clusters to recommend genres, and in clusters that do not have relatively specific tendencies, genre prediction has been performed by appropriately trimming genres that are not necessary for recommendation in order to improve performance.
We have performed various experiments using
our method on well-known movie datasets, and the results showed that clustering and sub-genre trimming worked, but the AVT was not that great.

%%%%%%%%%%%%%%%%%%%%%%%%%%%%%%%%%%%%%%%%%%
%% optional
%\supplementary{The following are available online at \linksupplementary{s1}, Figure S1: title, Table S1: title, Video S1: title.}


%%%%%%%%%%%%%%%%%%%%%%%%%%%%%%%%%%%%%%%%%%

% \acknowledgments{In this section you can acknowledge any support given which is not covered by the author contribution or funding sections. This may include administrative and technical support, or donations in kind (e.g., materials used for experiments).}

%%%%%%%%%%%%%%%%%%%%%%%%%%%%%%%%%%%%%%%%%%
%% Only for journal Encyclopedia
%\entrylink{The Link to this entry published on the encyclopedia platform.}

%%%%%%%%%%%%%%%%%%%%%%%%%%%%%%%%%%%%%%%%%%
%% Optional

%%%%%%%%%%%%%%%%%%%%%%%%%%%%%%%%%%%%%%%%%%
%% Optional

%%%%%%%%%%%%%%%%%%%%%%%%%%%%%%%%%%%%%%%%%%

\section*{References}
\end{document}